\documentclass[sigconf,balance=false]{acmart}
\usepackage{popets}
\setcopyright{popets}
\copyrightyear{2025}

\acmYear{YYYY}
\acmVolume{YYYY}
\acmNumber{X}
\acmDOI{XXXXXXX.XXXXXXX}
\acmISBN{}
\acmConference{Proceedings on Privacy Enhancing Technologies}
\usepackage{amsmath,amsfonts,amsthm}
\usepackage{mathtools}

\usepackage{soul,bm}
\usepackage{relsize}
\usepackage{comment}
\usepackage[symbol]{footmisc} 

\usepackage{graphicx}
\usepackage[justification=centering,font=small]{caption}
\usepackage{subfigure,subcaption,wrapfig}

\usepackage{tabularx}
\usepackage{booktabs}
\usepackage{threeparttable}
\usepackage{multirow}
\usepackage{makecell}

\usepackage{algorithmic}
\usepackage[ruled,vlined,linesnumbered]{algorithm2e}

\usepackage{tikz}
\usepackage{pgfplots,pgfplotstable}
\pgfplotsset{compat=1.7}

\newtheorem{definition}{\textbf{Definition}}
\newtheorem{theorem}{\textbf{Theorem}}
\newtheorem{lemma}{\textbf{Lemma}}
\newtheorem{example}{\textbf{Example}}

\begin{document}

\title{Locally Differentially Private Frequency Estimation via Joint Randomized Response}

\author{Ye Zheng}
\orcid{0000-0003-0623-9613}
\affiliation{%
  \institution{Rochester Institute of Technology}
  \city{} 
  \state{}
  \country{}}
\email{ye.zheng@mail.rit.edu}

\author{Shafizur Rahman Seeam}
\orcid{0000-0003-3350-0047}
\affiliation{
  \institution{Rochester Institute of Technology}
  \city{}
  \state{}
  \country{}}
\email{ss6365@rit.edu}

\author{Yidan Hu}
\orcid{0000-0002-9443-8411}
\affiliation{
  \institution{Rochester Institute of Technology}
  \city{}
  \state{}
  \country{}}
\email{yidan.hu@rit.edu}

\author{Rui Zhang}
\orcid{0000-0001-5230-5998}
\affiliation{
  \institution{University of Delaware}
  \city{}
  \state{}
  \country{}}
\email{ruizhang@udel.edu}

\author{Yanchao Zhang}
\orcid{0000-0001-5230-5998}
\affiliation{
  \institution{Arizona State University}
  \city{}
  \state{}
  \country{}}
\email{yczhang@asu.edu}


\renewcommand{\shortauthors}{Zheng et al.}

\begin{abstract}
Local Differential Privacy (LDP) has been widely recognized as a powerful tool for providing a 
strong theoretical guarantee of data privacy to data contributors against an untrusted data collector. 
Under a typical LDP scheme, each data contributor independently randomly perturbs their data before submitting them to the data collector, 
which in turn infers valuable statistics about the original data from received perturbed data. 
Common to existing LDP mechanisms is an inherent trade-off between the level of privacy protection and 
data utility in the sense that strong data privacy often comes at the cost of reduced data utility. 
Frequency estimation based on Randomized Response (RR) is a fundamental building block of many LDP mechanisms. 
In this paper, we propose a novel Joint Randomized Response (JRR) mechanism based on correlated data perturbations 
to achieve locally differentially private frequency estimation. 
JRR divides data contributors into disjoint groups of two members and lets those in the same group 
jointly perturb their binary data to improve frequency-estimation accuracy and achieve the same level of 
data privacy by hiding the group membership information in contrast to the classical RR mechanism. 
Theoretical analysis and detailed simulation studies using both real and synthetic datasets show that 
JRR achieves the same level of data privacy as the classical RR mechanism while improving the frequency-estimation 
accuracy in the overwhelming majority of the cases by up to two orders of magnitude.   
\end{abstract}

\keywords{local differential privacy, randomized response, frequency estimation}
\maketitle

\section{Introduction}
Differential privacy \cite{DworkAlg14} is widely considered as the \textit{de facto} framework for providing strong theoretical guarantee of data privacy. Recent years have also witnessed significant interests in developing data analysis techniques for ensuring differential privacy in the local setting, commonly referred to as Local Differential Privacy (LDP) \cite{DuchiLoc13}. A local differential privacy mechanism protects individual data contributors' data privacy against an untrusted data collector by having each data contributor randomly perturb their data value before submission while allowing the data collector to learn valuable statistics of the contributors' data. 
In addition to significant interests from academia, LDP techniques have seen growing adoption by industry for various data analysis applications. For example, Google has deployed RAPPOR \cite{ErlinRAP14} into Chrome to privately collect individual web browsing behavior. As another example, Apple adopts LDP algorithms~\cite{Apple17} in Safari for privacy-preserving collection of users' typing history to better understand user behaviors.

Significant efforts have been made to achieve a good utility-privacy trade-off for various data analysis tasks. In particular, all existing LDP mechanisms exhibit a natural trade-off between data privacy and data utility at the data collector because strong data privacy for individual data contributors often comes at the cost of reduced data utility \cite{DuchiLoc13,RoyCry20}.  
Therefore, a major focus of current research is to design LDP mechanisms that achieve higher data utility without sacrificing privacy guarantees for individual contributors.
For example, several recent proposals \cite{ErlinAmp19,BallePri19,BittaPro17,CheuDis19,MeehaPri22} show that it is possible to improve privacy protection while reducing the amount of noise needed by having an auxiliary server shuffle data contributors' perturbed data before sending them to the data collector. 
Other proposed approaches include parameter optimization \cite{KairoExt16,WangLoc17}, developing advanced encoding schemes \cite{BassiLoc15,ErlinRAP14,KairoDis16}, random perturbation schemes and estimators \cite{WangLoc17,MurakUti19}, interactive data collection schemes \cite{YePri19}, cryptography-assisted solutions \cite{RoyCry20}, post-processing techniques \cite{WangLocally20, fang2023locally}, etc.

Frequency estimation is a classical data analysis problem in which the data collector aims to learn the number (or ratio) of data contributors with a certain attribute or possessing a particular data value. Randomized Response (RR) \cite{WarneRan65} is the first known and most classical LDP protocol for frequency estimation. Since frequency estimation is used in many other data analysis tasks such as heavy hitter estimation, mean value estimation, and range queries, RR is also widely used as a fundamental building block in many LDP mechanisms for these tasks \cite{BassiLoc15,WangLoc18,WangLoc21,WangCol19,CormoAns19,WangAns19}. Common to these solutions is that every data contributor independently perturbs his/her data before submitting them to the data collector. \emph{An open question is whether it is possible to improve data utility of RR-based LDP mechanisms without loss of LDP guarantees by introducing correlations among the random perturbations performed by different data contributors}.

In this paper, we make the first attempt to explore correlated random perturbations for frequency estimation to improve data utility without any sacrifice of LDP guarantees. We observe that it is possible to achieve much higher data utility in terms of estimation accuracy at the data collector by randomly dividing data contributors into disjoint groups of two and introducing carefully crafted correlations to each group's random perturbations. At the same time, no additional information can be inferred as long as the group membership is kept secret from the data collector. Based on these observations, we introduce a novel \textbf{Joint Randomized Response (JRR)} mechanism for locally differentially private frequency estimation. 
By carefully tuning the parameters, JRR can achieve significantly higher data utility in the overwhelming majority of cases while offering the same level of LDP protection as the classical RR mechanism. 

Our contributions in this paper can be summarized as follows.
\begin{itemize}

    \item We are the first to explore correlated random perturbations for frequency estimation to improve data utility at the data collector without sacrificing LDP guarantee for individual data contributors. 

    \item We introduce a general Joint Randomized Response (JRR) mechanism that achieves the same level of LDP protection as the classical RR mechanism, while improving the data utility in an overwhelming majority of the cases, especially for a large number of data contributors.
    \item We present a practical instantiation of JRR by utilizing a non-colluding auxiliary server. 
    \item We thoroughly evaluate JRR via a combination of theoretical analysis and detailed simulation studies using both real and synthetic datasets. Our results show that JRR outperforms the classical RR mechanism for over 97\% of the possible frequencies and improve the estimation accuracy by as much as two orders of magnitude.
    
\end{itemize}
The rest of the paper is structured as follows. Section~\ref{Sec:problem} presents the problem formulation and reviews LDP and the RR mechanism. Section~\ref{sec:impact} uses two examples to demonstrate the impact of correlated random perturbations.
Section~\ref{Sec:Theory} introduces a general JRR mechanism, its performance analysis, and a practical instantiation. Section~\ref{Sec:Eval} evaluates the performance of JRR. Section~\ref{Sec:Related} discusses related work. Section~\ref{Sec:Conl} concludes this paper and points out several future research directions.

\section{Preliminaries} \label{Sec:problem}

In this section, we formulate the problem and then reviews LDP and the RR mechanism.

\subsection{Problem Formulation}

We consider a system consisting of a data collector and a set of data contributors $\mathcal{U}=\{u_1,u_2,\cdots,u_n\}$. Each contributor has a binary value $x_i\in D=\{0,1\}$, and the data collector wants to learn the number of data contributors having value $v$ for each $v\in D$. Data contributors are concerned about their data privacy. As a result, instead of submitting the original value $x_i$, each contributor $u_i$ randomly perturbs his/her value using a random perturbation mechanism $\mathcal{M}$ and submits the perturbed value $y_i=\mathcal{M}(x_i)$ 
to the data collector. After receiving the perturbed data from $n$ contributors, the data collector estimates the number of data contributors having value $v$ for each $v\in D$.

We assume the data collector is honest but curious, meaning it faithfully carries out system operations but is interested in inferring the original data values of the contributors. Specifically, we assume the data collector will not register or create fake contributor accounts to participate in data collection, as doing so would risk damaging its business reputation if detected. Moreover, we assume that normal data contributors are concerned about their data privacy and will not disclose their original data values to the data collector. Even if a few data contributors collude with the data collector, we assume that the number of such contributors does not exceed a predefined threshold $M$, e.g., a small fraction of all the contributors.

We seek to design a locally differentially private frequency estimation scheme that enables the data collector to estimate $n_v$ with high accuracy while providing individual contributors with the same $\varepsilon$-LDP guarantee as the classical RR mechanism.

\subsection{Local Differential Privacy}

Local Differential Privacy (LDP) is considered a gold standard for privacy-preserving data collection against an untrusted data collector. 
It requires a perturbation mechanism that provides enough randomness to the private data.

\begin{definition}[\textbf{Local Differential Privacy}]\label{def:LDP}
A randomized mechanism $\mathcal{M}: \mathcal{X}\to\mathsf{Range}(\mathcal{M})$ satisfies $\varepsilon$-LDP if
\begin{equation}\label{eq:LDP}
    \frac{\Pr[\mathcal{M}(x)=y]}{\Pr[\mathcal{M}(x')=y]}\leq e^{\varepsilon},
\end{equation}
for any inputs $x,x'\in \mathcal{X}$ and any output $y\in \mathsf{Range}(\mathcal{M})$, where $\mathsf{Range}(\mathcal{M})$ is the output range of $\mathcal{M}$.
\end{definition}

Here $\varepsilon$ is a parameter controlling the level of privacy protection commonly referred to as 
\emph{privacy budget}. 
The smaller the $\varepsilon$, the stronger the privacy protection, and vice versa. 
Intuitively, $\varepsilon$-LDP means that by observing the output $y$, 
the data collector cannot infer whether the input is $x$ or $x'$ with high confidence, 
which provides contributors submitting sensitive data with plausible deniability.

\subsection{Review of Randomized Response}
Randomized Response (RR)~\cite{WarneRan65} was originally proposed to provide plausible deniability to 
interviewees answering a sensitive boolean question in a survey. 
Under RR, each interviewee reports the answer truthfully with probability $p$ (the opposite answer with $q=1-p$). 
RR mechanism satisfies $\varepsilon$-LDP if $p\leq e^{\varepsilon}/(1+e^{\varepsilon})$.

Assume that the total number of data contributors is $n$ and that $n_v$ contributors have value $v$ for each $v\in D$. Suppose that the data collector receives $I_v$ 
perturbed value $v$. 
The data collector estimates the number of data contributors having value $v$ as
\begin{equation}\label{eq:est_rr}
   \hat{n}_v=\frac{I_v-nq}{p-q},
\end{equation}
which is an unbiased estimator of $n_v$~\cite{WangLoc17,WarneRan65}.

The data utility of RR is commonly measured by the variance of the unbiased estimator 
$\hat{n}_v$, which is given by
\begin{equation}\label{eq:var_rr}
    \mathrm{Var}[\hat{n}_v]=\frac{\mathrm{Var}[I_v]}{(p-q)^2}=\frac{npq}{(p-q)^2}.
\end{equation}

\section{Impact of Correlation Among Data Contributors}\label{sec:impact}

In this section, we discuss the potential impact of introducing correlations among the random perturbations performed by different data contributors on data privacy and data utility through examples.

The data utility of LDP protocols such as \cite{WangLoc17,WarneRan65}, is commonly measured by the variance of the estimator of the value of interest. A smaller variance indicates higher data utility. Traditional LDP protocols involve each contributor independently perturbing their data. Consequently, the estimator of an LDP protocol is reduced to the sum of the individual contributors' reported values, and its variance is proportional to the sum of the variances of individual reported values. 

We find that if multiple contributors jointly perturb their data, the variance of the estimator also depends on the covariance of the jointly perturbed values. By carefully designing the joint perturbation to introduce a negative covariance, it is possible to achieve higher data utility. In what follows, we use two concrete examples to illustrate this finding.

\begin{example}\label{ex:RR}
(Independent perturbation)
Suppose that there are two data contributors, $u_1$ and $u_2$ with values $x_1=1$ and $x_2=1$, respectively. Each contributor independently perturbs their value using RR with $p=0.8$. 
Let $T_j$ be the indicator of reporting truthfulness of $u_j$, 
i.e., $T_j=1$ if $y_j=x_j$ and $0$ otherwise. We have
\begin{equation}\label{eq:PDRR}
T_j=\begin{cases}
1& \text{with probability $p=0.8$},\\
0& \text{with probability $q=0.2$}.\\
\end{cases}
\end{equation}
\end{example}

\textbf{Estimation of \boldmath{$n_1$}:} 
Assume that the data collector has received $I_1$ perturbed values of $1$. According to Eq.~(\ref{eq:est_rr}), the data collector can estimate $n_1$ as
\begin{equation}\label{eq:EstN1}
\hat{n}_1=\frac{(I_1-2\times 0.2)}{0.6}.
\end{equation}

\textbf{Data privacy:} Since $p/q=0.8/0.2=4$, the RR mechanism in the above example satisfies $\ln 4$-LDP.

\textbf{Data utility:} According to Eq.~(\ref{eq:var_rr}), the variance of $\hat{n}_1$ can be computed as 
\begin{equation}\label{eq:mse_rr_eg}
\begin{split}
\mathrm{Var}[\hat{n}_1]&=\frac{npq}{(2p-1)^2}=\frac{2\cdot 0.8\cdot 0.2}{(2\cdot 0.8-1)^2}=0.89.
\end{split}
\end{equation}

\begin{table}[t]
\small
    \centering
    \caption{Joint probability distribution in Example~\ref{ex:JRR}. } 
    \label{tab:example_joint_prob}
    \begin{tabular}{ ccc }
        \toprule
        & $T_{1}=1$    & $T_{1}=0$    \\
        \midrule
        $T_{2}=1$ & $0.61$ & $0.19$ \\
        \vspace{-0.6em} \\
            $T_{2}=0$ & $0.19$ & $0.01$ \\
        \bottomrule
    \end{tabular}
\end{table}

\begin{example}\label{ex:JRR}
(Correlated perturbation)
Consider the same two contributors in Example \ref{ex:RR}. 
Let $T_j$ be a binary indicator of a random variable for whether a data contributor $u_j$ reports truthfully, 
i.e., $T_j=1$ if $y_j=x_j$ and 0 otherwise. The two contributors jointly perturb their data according to the joint probability distribution shown in Table~\ref{tab:example_joint_prob}.
\end{example}

\textbf{Estimation of \boldmath{$n_1$}:} 
It is easy to see that the marginal probability distribution of both $T_1$ and $T_2$ in Table~\ref{tab:example_joint_prob} is
the same as the probability distribution of $T_j$ in Example \ref{ex:RR}. 
Define $Y_j$ to be the indicator random variable for data contributor $u_j$ reports a perturbed value $y_j=1$ for all $1\leq j\leq 2$. There are two cases. First, if $x_j=0$, we have $\mathrm{Pr}[Y_j=1|x_j=0]=\mathrm{Pr}[T_j=0]=0.2$. Second, if $x_j=1$, we have $\mathrm{Pr}[Y_j=1|x_j=1]=\mathrm{Pr}[T_j=1]=0.8$. Let $I_1$ be the random variable for the number of contributors reporting a perturbed value $1$. We have $I_1=Y_1+Y_2$. Taking the expectation on both sides, we have
\begin{equation}
\begin{split}
\mathrm{E}[I_1]&=\mathrm{E}[\sum_{j=1}^{2}Y_j]=\sum_{j=1}^{2}\mathrm{E}[Y_j]=\sum_{j=1}^{2}\mathrm{Pr}[Y_j=1]\\
&=n_1\cdot \mathrm{Pr}[T_j=1] + (2-n_1)\cdot 1\cdot \mathrm{Pr}[T_j=0]\\
&=0.8n_1+0.2\cdot(2-n_1)\\
&=0.4+0.6n_1.
\end{split}
\end{equation}

The data collector can estimate $n_1$ as $\hat{n}_1=(I_1-0.4)/{0.6}$, which is an unbiased estimator of $n_1$ and also identical to Eq.~(\ref{eq:EstN1}) in Example~\ref{ex:RR}.

\textbf{Data privacy:} Since the marginal probability distribution of both $T_1$ and $T_2$ in Table~\ref{tab:example_joint_prob} is the same as the one in Example \ref{ex:RR}, those marginal probability distributions also satisfy $\ln 4$-LDP.
However, it does not indicate that each contributor can enjoy the same level of $\ln 4$-LDP as in Example \ref{ex:RR}. In fact, the introduction of correlation among
different contributors will inevitably reduce privacy guarantee for individual contributors.
We postpone the discussion of the potential privacy leakage from the correlation between two contributors in the same group to Section \ref{subsec:CRRM_privacy}.

\textbf{Data utility:} 
The variance of the unbiased estimator is
\begin{equation}\label{eq:mse_eg}
\begin{split}
    \mathrm{Var}[\hat{n}_1]&=\frac{\mathrm{Var}[I_1]}{0.36}
    =\frac{25}{9} \mathrm{Var}[Y_1+Y_2]\\
    &=\frac{25}{9} (\mathrm{Var}[Y_1]+\mathrm{Var}[Y_2]+2\mathrm{Cov}[Y_1,Y_2]),
\end{split}
\end{equation}
where $\mathrm{Cov}[Y_1,Y_2]$ is the covariance between $Y_1$ and $Y_2$.

First, $\mathrm{Var}[Y_1]$ and $\mathrm{Var}[Y_2]$ are the same due to the same marginal distribution. Moreover, since both contributors have the same original value of $1$, we have $\mathrm{E}[Y_j]=\mathrm{Pr}[Y_j=1]=\mathrm{Pr}[T_j=1]=0.8$ and $\mathrm{E}[Y^2_j]=\mathrm{Pr}[Y_j=1]=0.8$. It follows that
\begin{equation}\label{eq:var_eg}
\begin{split}
     \mathrm{Var}[Y_1]+\mathrm{Var}[Y_2]&=2\mathrm{Var}[Y_1]=2(\mathrm{E}[Y^2_1]-\mathrm{E}^2[Y_1])\\
     &=2\cdot (0.8-0.8^2)=0.32.
\end{split}
\end{equation}

We now compute $\mathrm{Cov}[Y_1, Y_2]$. Since $x_1 = x_2 = 1$, we have
    \begin{equation*}
        \begin{split}
            \mathrm{E}[Y_1Y_2] =& 0 \times \Pr[T_{1}=0, T_{2}=0] + \\
            & 1 \times \Pr[T_{1}=1, T_{2}=1]=0.61,
        \end{split}
    \end{equation*}
and it follows that
\begin{equation}\label{eq:cov_eg}
\begin{split}
        \mathrm{Cov}[Y_1, Y_2] &= \mathrm{E}[Y_1Y_2]-\mathrm{E}[Y_1]\mathrm{E}[Y_2]\\
    &=0.61-0.8\cdot 0.8=-0.03.
\end{split}
\end{equation}
Substitute Eqs.~(\ref{eq:var_eg}) and (\ref{eq:cov_eg}) into Eq.~(\ref{eq:mse_eg}), we have
\begin{equation}
     \mathrm{Var}[\hat{n}_1]=\frac{1}{0.36}(0.32 +2\times(-0.03))\approx 0.72,
\end{equation}
which is smaller than the $\mathrm{Var}[\hat{n}_1]$ of 0.89 in Example \ref{ex:RR}.

From the above two examples, we can see that it is possible to improve data utility, i.e., reduce the variance of the estimator, through the introduction of a negative correlation between $Y_1$ and $Y_2$ via joint perturbation of two contributors. 
Theoretical analysis of generalizing $n$ and $x_i$ in the above examples will be presented in the next section.
Meanwhile, several key questions must be answered to fully exploit the potential of joint random perturbation. 

\begin{enumerate}

   \item   How can we generalize the above joint perturbation mechanism given in Table~\ref{tab:example_joint_prob}?
    \item 
    Can the joint perturbation mechanism provide the same level of data privacy as RR? If so, under what condition? In particular, is it possible for the data collector to infer additional information about a target contributor's value by exploiting the correlations among different data contributors? 
    \item How can we quantify the data utility of a joint perturbation mechanism?
    \item 
    How can we optimize the joint perturbation mechanism to maximize the data utility while guaranteeing the same level of data privacy as RR?
    \item  How can we design a practical joint perturbation mechanism?
\end{enumerate}

We provide answers to these questions in the next section.

\section{Joint Randomized Response} \label{Sec:Theory}
This section first introduces a general joint randomized
response (JRR) mechanism as a generalization of
the classical RR mechanism. We then generalize the definition of LDP to ensure that JRR can provide the same level of privacy protection as the classical RR. We quantify the data utility of JRR in Section~\ref{subsec:CRRM_utility}. Section~\ref{subsec:CRRM_opt} presents a heuristic algorithm to choose its parameters for maximized data utility given the desirable level of privacy protection. Finally, we present two practical instantiations of the JRR mechanism.

\subsection{A General JRR Mechanism}\label{subsec:CRRM}

Assume there are $n$ contributors, $\mathcal{U} = \{u_1,\dots, u_n\}$ each having a binary value and $n$ is an even number. We first divide the $n$ contributors into $n/2$ disjoint groups of two $G_1,\dots, G_{n/2}$ uniformly at random.
Without loss of generality, assume that each group $G_i$ consists of contributors $u_{2i-1}$ and $u_{2i}$ for all $1\leq i\leq {n}/{2}$. Each group $G_i$ of two contributors jointly perturb their values according to the joint probability distribution shown in Table~\ref{tab:jointPro_2}, where $0.5< p\leq 1$, $q=1-p$, and $1-{1}/{p}\leq \rho\leq 1$ are system parameters. 
In particular, each contributor $u_j, 1\leq j\leq n,$ reports $y_j=x_j$ if $T_j=1$ and $1-x_j$ if $T_j=0$.

\subsubsection{Properties of JRR}
The general JRR mechanism has several key properties, which are summarized as follows.

First, the marginal probability distribution of every $T_j$ $(1\leq j\leq n)$ is identical. In particular, it is easy to verify that
\begin{equation}
    \begin{split}
        \Pr[T_j=1]=& p^2+\rho p q+(1-\rho) pq=p \\
        \Pr[T_j=0]=&(1-\rho) pq+q^2+\rho pq=q.
    \end{split}
\end{equation}
This means that each data contributor reports their value truthfully with probability $p$ and untruthfully with probability $q=1-p$, which aligns  with the classical RR mechanism with parameter $p$.

\begin{table}[t]
\small
    \centering
    \caption{Joint reporting probability in JRR.} \label{tab:jointPro_2}
    \begin{tabular}{ ccc }
        \toprule
        & $T_{2i-1}=1$    & $T_{2i-1}=0$    \\
        \midrule
        $T_{2i}=1$ & $p^2+\rho pq$ & $(1-\rho) pq$ \\
        \vspace{-0.6em} \\
        $T_{2i}=0$ & $(1-\rho) pq$ & $q^2+\rho pq$ \\
        \bottomrule
    \end{tabular}
\end{table}

Second, parameter $\rho$ is the correlation coefficient between random variables $T_{2i-1}$ and $T_{2i}$. Specifically, the correlation coefficient between random variables $T_{2i-1}$ and $T_{2i}$ is given by
\begin{equation}\label{eq:def_rho}
    \frac{\mathrm{Cov}[T_{2i-1},T_{2i}]}{\sigma_{1}\sigma_{2}}=\frac{\mathrm{E}[T_{2i-1}T_{2i}]-\mathrm{E}[T_{2i-1}]\mathrm{E}[T_{2i}]}{\sigma_{1}\sigma_{2}},
\end{equation}
where $\mathrm{Cov}[T_{2i-1},T_{2i}]$ is the covariance of $T_{2i-1}$ and $T_{2i}$, and $\sigma_{1}$ and $\sigma_{2}$ are the standard deviation of $T_{2i-1}$ and $T_{2i}$, respectively.
According to the joint probability distribution in Table.~\ref{tab:jointPro_2}, we have
\begin{equation}\label{eq:E_T1T2}
\begin{split}
\mathrm{E}[T_{2i-1}T_{2i}]&=\mathrm{Pr}[T_{2i-1} T_{2i}=1]\cdot 1+\mathrm{Pr}[T_{2i-1}T_{2i}=0]\cdot 0\\
&=\mathrm{Pr}[T_{2i-1}=1,T_{2i}=1]\cdot 1\\
&=\rho pq+p^2.
\end{split}
\end{equation}
We can also compute
\begin{equation}\label{eq:E_Tj}
    \mathrm{E}[T_{j}]=\mathrm{Pr}[T_{j}=1]\cdot 1+\mathrm{Pr}[T_{j}=0]\cdot 0=p
\end{equation}
 and
\begin{equation}\label{eq:sigma_Tj}
\begin{split}
     \sigma^2_j&=\mathrm{E}[T^2_j]-\mathrm{E}^2[T_j]\\
     &=\mathrm{Pr}[T^2_j=1]\cdot 1+ \mathrm{Pr}[T^2_j=0]\cdot 0\\
     &\quad-(\mathrm{Pr}[T_j=1]\cdot 1+ \mathrm{Pr}[T_j=0]\cdot 0)^2\\
     &=\mathrm{Pr}[T^2_j=1]-\mathrm{Pr}[T_j=1]^2\\
     &=\mathrm{Pr}[T_j=1]-\mathrm{Pr}[T_j=1]^2\\
     &=p-p^2
\end{split}
\end{equation}
for all $1\leq j\leq n$.

Substituting Eqs.~(\ref{eq:E_T1T2}) to~(\ref{eq:sigma_Tj}) into Eq.~(\ref{eq:def_rho}), we get the correlation coefficient between $T_{2i-1}$ and $T_{2i}$ as
\begin{equation}
\begin{split}
    \frac{\mathrm{Cov}[T_{2i-1},T_{2i}]}{\sigma_{1}\sigma_{2}}&=\frac{\rho pq+p^2-p\cdot p}{p-p^2}\\
    &=\rho.
\end{split}
\end{equation}
Note that $\rho$ must not be smaller than $1-{1}/{p}$ to ensure that every probability value in Table~\ref{tab:jointPro_2} is non-negative.

Third, the classical RR mechanism is a special case of
the general JRR mechanism. In particular, when $\rho = 0$, $T_{2i-1}$ and $T_{2i}$ are independent, and the JRR mechanism is equivalent to the case of each two in each perturbs his/her data value using the RR independently.

Last but not least, the data collector can estimate $n_v$ using the same estimator as in the classical RR. Specifically, assume that the data collector receives $I_v$ values of $v$ for each  $v\in D$, it can estimate the number of  contributors having true value $v$ as
\begin{equation}\label{eq:JRRestimator}
\hat{n}_v=\frac{I_v-nq}{p-q},
\end{equation}

\begin{theorem}\label{thm:uEst_JRR}
    The estimator in Eq.(\ref{eq:JRRestimator}) is unbiased.
\end{theorem}
We give the proof in Appendix~\ref{appendix:unbiased_jrr}.

\subsection{Data Privacy Analysis}\label{subsec:CRRM_privacy}
Assume that the data collector wants to infer a target contributor $u_i$'s value $x_i$. Let $\mathcal{C}$ be the set of contributors who collude with the data collector so that the data collector knows whether each contributor in $\mathcal{C}$ reports truthfully. In other words, besides all perturbed values $y_1,\cdots,y_n$, 
the data collector also knows whether each colluding contributor reports truthfully or not, which we denote by $\mathcal{T}_c=\{T_j|j\in\mathcal{C}\}$. Under the uniformly random grouping of JRR, the group peer of $u_i$, say $u_j$, may be a colluding contributor in $\mathcal{C}$. If this happens, even if the data contributor does not know who the group peer of contributor $u_i$ is, the correlation between the two contributors' perturbation would still inevitably reduce the data privacy of the targeted contributor $u_i$.

We therefore extend the definition of LDP to measure the individual privacy provided by the JRR scheme. Specifically, the following theorem shows that the JRR scheme can offer individual contributors a form of data privacy similar to $\varepsilon$-LDP.

\begin{theorem}\label{theo:privacy_level}
Assume that there is a set of contributors $\mathcal{C}$ whose reporting truthfulness $\mathcal{T}_c$ is known to the adversary.
For any contributor $u_i$,  the JRR mechanism $\mathcal{M}$ satisfies
\begin{equation}\label{eq:privacy}
\begin{split}
    &\frac{\mathrm{Pr}[\mathcal{M}(x_i)=y_i|\mathcal{T}_c]}{\mathrm{Pr}[\mathcal{M}(x'_i)=y_i|\mathcal{T}_c]}\leq e^\varepsilon\\
\end{split}
\end{equation}
for any pair of inputs $x_i,x'_i\in D$ and any output $y_i\in \mathsf{Range}(\mathcal{M})$, where
\begin{equation}\label{eq:epsilon}
        \varepsilon= \ln \frac{mp_{\max}+(n-m-1)p}{mp_{\min}+(n-m-1)q},
\end{equation}
with $ p_{\max}=\max\{(1-\rho)p,p+\rho q\}$, $p_{\min}=\min\{(1-\rho)q,q+\rho p\}$, $m=|\mathcal{T}_c|$, and $0\leq m\leq n-1$.
\end{theorem}
The detailed proof is provided in Appendix~\ref{appendix:jrr_privacy}.

\subsection{Data Utility Analysis} \label{subsec:CRRM_utility}

\begin{theorem}\label{thm:variance}
    Assume that $n$ contributors are divided into
$n/2$ groups uniformly at random. The variance of the estimator $\hat{n}_v$ given in Eq. (\ref{eq:JRRestimator}) by JRR is
    \begin{equation}\label{eq:variance_2u}
        \mathrm{Var}[\hat{n}_v]=
        \frac{pq}{(p-q)^2}\cdot \left(n+\frac{\rho\left((2n_1-n)^2-n\right)}{n-1}\right),
    \end{equation}
where $n_v$ is the number of contributors with a value of $v$.
\end{theorem}

The proof is given in Appendix~\ref{appendix:jrr_utility}.

We can see that $\mathrm{Var}[\hat{n}_v]=
        {npq}/{(p-q)^2}$ when $\rho=0$, which is the same as that of RR.
Moreover,  $\mathrm{Var}[\hat{n}_v]<
        {npq}/{(p-q)^2}$, i.e., smaller than that of RR, if $\rho((2n_1-n)^2-n)<0$, which provides opportunities to achieve higher data utility than RR.

\subsection{Choice of $p$ and $\rho$} \label{subsec:CRRM_opt}
In this subsection, we show how to choose parameters $p$ and $\rho$ to achieve high data utility at the data collector under a given data privacy requirement.

\subsubsection{An Optimization Problem Formulation}
Assume that we want to provide the same level of privacy guarantee with an RR scheme that satisfies $\varepsilon$-LDP. We need to choose parameters $p$ and $\rho$ that satisfies Ineq.~(\ref{eq:privacy}) in Theorem~\ref{theo:privacy_level}. 
One challenge is that Ineq.~(\ref{eq:privacy}) involves the parameter $m$, i.e., the number of contributors colluding with the data collector, 
which is often unknown in practice. 
Fortunately, we find that for any pair of $(p,\rho)$, given the privacy budget $\varepsilon$, if $m_1$ satisfies Ineq.~(\ref{eq:privacy}), so does $m_2$ for all $ m_2\leq m_1$, because $f(m)=\frac{mp_{\max}+(n-m-1)p}{mp_{\min}+(n-m-1)q}$ is monotonically increasing with respect to $m$. 
The detailed proof is given in Appendix~\ref{appendix:proof:m_M}. 
Assume that there could be at most $M$ data contributors colluding with the data collector.
We can then replace $m$ in Ineq.~(\ref{eq:privacy}) by $M$ when choosing $p$ and $\rho$.

Let $h(p,\rho)= \mathrm{Var}[\hat{n}_1]$ as given in Eq.~(\ref{eq:variance_2u}). We can formulate the choice of $p$ and $\rho$ as the following optimization problem, which seeks to minimize the objective function $h(p,\rho)$ while satisfying the privacy constraint and the domain constraint of $\rho$ and $p$.

\begin{equation}\label{eq:opt_pro}
 \begin{split}
\min& \quad h(p,\rho)\\
\text{s.t.}
&\quad \frac{Mp_{\max}+p(n-M-1)}{Mp_{\min}+q(n-M-1)}\leq e^\varepsilon,\\
&\quad1-1/p\leq \rho\leq 1,\\
&\quad0.5 < p\leq 1,
\end{split}
\end{equation}
where $ p_{\max}=\max\{(1-\rho)p,p+\rho q\}$ and $p_{\min}=\min\{(1-\rho)q,q+\rho p\}$.

Unfortunately, since the objective function $h(p,\rho)$ involves $n_1$ that we want to estimate, the above optimization problem cannot be directly solved without knowing $n_1$ in advance. However, certain properties of the objective function $h(p,\rho)$ and the constraints make it possible to design an effective heuristic to choose $p$ and $\rho$ that can yield good performance in the majority of the cases. 

Specifically, we have the following three lemmas,  with proofs in Appendix~\ref{appendix:lemma:1}, Appendix~\ref{appendix:lemma:2}, and Appendix~\ref{appendix:lemma:3}, respectively.
\begin{lemma}\label{lemma:1}
    For any $n$, $n_1 > 0$, the following inequality holds:
\begin{equation}
(2n_1-n)^2-n\begin{cases}\geq 0&\text{if $\frac{n_1}{n}\in[0,\frac{1}{2}-\frac{1}{2\sqrt{n}}]\bigcup (\frac{1}{2}+\frac{1}{2\sqrt{n}},1)$},
\\
< 0 & \text{if } \frac{n_1}{n}\in[\frac{1}{2}-\frac{1}{2\sqrt{n}},\frac{1}{2}+\frac{1}{2\sqrt{n}}].
\end{cases}
\end{equation}
\end{lemma}

\begin{lemma}\label{lemma:2}
For any $n\geq 2, \rho\in[-1,1]$ and $0\leq n_1\leq n$,
\begin{equation}
n+\frac{\rho((2n_1-n)^2-n)}{n-1}>0.
\end{equation}
\end{lemma}

\begin{lemma}\label{lemma:3}
${pq}/{(p-q)^2}$ is monotonically decreasing with respect to $p\in (0.5,1]$.
\end{lemma}

We then have the following theorem regarding the monotonicity of the objective function $h(p,\rho)$.
\begin{theorem}\label{thm:mono}
The objective function $h(p,\rho)$ is
\begin{itemize}
    \item monotonically increasing with respect to $\rho$ if $n_1/n\leq 1/2-1/2\sqrt{n}$ or $ n_1/n\geq 1/2+1/2\sqrt{n}$ and monotonically decreasing with respect to $\rho$ if $1/2-1/2\sqrt{n}<n_1/n<  1/2+1/2\sqrt{n}$,
    \item monotonically decreasing with respect to $p\in(0.5,1]$.
\end{itemize}
\end{theorem}
The proof uses the results from Lemmas~\ref{lemma:1} to \ref{lemma:3}, which is quite straightforward and given in Appendix~\ref{appendix:thm:mono}. 

\subsubsection{A Heuristic Algorithm for Selecting $\rho$ and $p$}

We now introduce a heuristic to choose $\rho$ and $p$ that can yield good performance in most cases by exploiting the monotonicity of  $h(p,\rho)$. First, we assume that the data collector would collude with at most $m=M$ contributors, where $M$ is a system parameter indicating the data collector's inference ability. 

According to Theorem~\ref{thm:mono}, $h(p,\rho)$ is monotonically increasing with respect to $\rho$ if $n_1/n\in [0, 1/2-1/2\sqrt{n}]\bigcup[ 1/2+1/2\sqrt{n},1]$ and monotonically decreasing with respect to $\rho$ if $n_1/n\in [1/2-1/2\sqrt{n},1/2+1/2\sqrt{n}]$.
We notice that the size of the range $[1/2-1/2\sqrt{n},1/2+1/2\sqrt{n}]$ is $1/\sqrt{n}$, which is relatively small for large $n$ and inversely proportional to $\sqrt{n}$. For example, when $n=100$ and $10,000$,  the ranges in which $h(p,\rho)$ is monotonically increasing with respect to $\rho$ are  $[0.45,0.55]$ and $[0.495,0.505]$, respectively. 
It is therefore reasonable to assume that $h(p,\rho)$ is monotonically increasing with respect to $\rho$ when $n$ is large in practice.

Assume that $n_1/n\notin [1/2-1/2\sqrt{n},1/2+1/2\sqrt{n}]$. The objective function $h(p,\rho)$ is then monotonically increasing with respect to $\rho$ and monotonically decreasing with respect to $p$ according to Theorem ~\ref{thm:mono}. Let $R\in\mathbb{R}^2$ be the feasible domain of $(p,\rho)$. We can define a partial ordering relationship among different pairs of $(p,\rho)$s. In particular, for any two pairs of $(p,\rho)$ and $(p',\rho')$, we have  $h(p,\rho)\leq h(p',\rho')$ if $p\geq p'$ and $\rho\leq \rho'$.
Moreover, let $\mathsf{range}(p)$ be the feasible range of $\rho$ for a given $p$. The optimal choice of the parameters must be $(p, \min(\mathsf{range}(p)))$ for some $p$.

Based on the above observation, our key idea is to first find a maximal feasible $p$ and then find the corresponding minimal feasible $\rho$. Let $\triangle p$ and $\triangle \rho$ be two small constants, e.g., $0.0001$. Algorithm~\ref{al:heu} provides the detailed procedure. Specifically, we initialize $p$ to $\frac{e^\varepsilon}{1+e^\varepsilon}-\triangle p$ (Line~1). We intentionally skip $p=\frac{e^\varepsilon}{1+e^\varepsilon}$ because if $p=\frac{e^\varepsilon}{1+e^\varepsilon}$ then $\rho$ would be zero and JRR would be equivalent to RR. We then search for a feasible pair of $p$ and $\rho$ using two nested loops. For each given $p$, we initialize $\rho$ to $1-\frac{1}{p}$ (Line~3) and then iteratively check whether the current $(p,\rho)$ satisfy all the constraints in Eq.~\ref{eq:opt_pro} (Lines 3-9). If so, we output $(p,\rho)$. Otherwise, we increase $\rho$ by $\triangle \rho$ until $\rho=1$, in which case a new iteration starts.
Algorithm~\ref{algo:search_rho} returns "Null" for completeness.
In practice, Algorithm~1 always returns a feasible pair $(\rho,p)$ in the worst case. The reason is that when $\rho =0$, we have $p_{\max}=p$ and $p_{\min}=q$, so the first constraint is simplified to $\frac{p}{q}\leq e^\varepsilon$. In this case, any $p\leq \frac{e^\varepsilon}{1+e^\varepsilon}$ with $\rho=0$ is always a feasible pair. The search space for $p$ and $\rho$ is $[0.5, \frac{e^\epsilon}{1+e^\epsilon})$ and $[1-\frac{1}{p},1]$, respectively, and the computational complexity of Algorithm~1 is proportional to the area of search space $\frac{1}{p}\cdot(\frac{e^{\varepsilon}}{1+e^{\varepsilon}}-0.5)$ and inversely proportional to both $\Delta p$ and $\Delta \rho$.

\begin{algorithm}[t]
    \small
    \SetAlgoLined
    \SetKwInOut{Input}{input}
    \SetKwInOut{Output}{output}
    \caption{Search for $(\rho,p)$ }\label{al:heu}
    \label{algo:search_rho}
    \Input{ $n$, $\varepsilon$, $M$, $\triangle \rho$, and $\triangle p$}
    \Output{$\rho$ and $p$}
    $p \leftarrow \frac{e^{\varepsilon}}{1+e^{\varepsilon}}-\triangle p$\;

    \While{$p >0.5$}
    {
        $\rho\leftarrow 1-\frac{1}{p}$\;

        \While{$\rho \leq 1$}
        {
            \If{$(p,\rho)$ satisfies all constraints in Eq.~(\ref{eq:opt_pro})}
            {
                
                \Return $p$ and $\rho$\;
            }
            $\rho\leftarrow \rho+\triangle\rho$\; 
        }
        $p\leftarrow p-\triangle p$\;
    }
    \Return \text{Null}\;
\end{algorithm}

\subsection{Practical Instantiations of JRR} \label{Sec:Practice}

In this subsection, we present two practical instantiations of the JRR mechanism for completeness. The first instantiation is highly efficient but requires an auxiliary non-colluding server to facilitate random grouping and joint random perturbation, as described in Table~\ref{tab:jointPro_2}. The second instantiation, in contrast, leverages multi-party computation techniques (MPC) to eliminate the need for a non-colluding server but comes at the cost of increased computational and communication overhead. While our proposed instantiations provide practical implementations, they are not necessarily optimal. There remains significant potential for further refinement and efficiency improvements, which we leave as future work.

\subsubsection{A Non-colluding Server-Based Instantiation}
Recall that the JRR mechanism relies on the two key assumptions: (1) all $n$ contributors are divided into $n/2$ groups uniformly at random,  and (2) the data collector is unaware of the group membership. 
Our first instantiation leverages an auxiliary non-colluding server to satisfy these requirements. Notably, similar non-colluding servers have been employed in recent works such as \cite{BittaPro17,ErlinAmp19,BallePri19,CheuDis19}. Moreover, various approaches have been proposed to enforce non-collusion, as discussed in \cite{DBLP:conf/eurocrypt/DworkKMMN06,DBLP:journals/cacm/Chaum81,DBLP:journals/popets/KwonLDF16}.

\textbf{Random grouping.} The auxiliary server divide 
$n$ contributors into $n/2$ disjoint groups of two uniformly at random. 
Without loss of generality, assume that the $i$-th group $G_i$ consists of contributors $u_{2i-1}$ and $u_{2i}$ for all $1\leq i\leq \frac{n}{2}$. For each group $G_i$, the server generates $R_{2i-1}\in\{1,-1\}$ uniformly at random and computes $R_{2i}=-R_{2i-1}$. It then sends $R_{2i-1}$ to $u_{2i-1}$ and $R_{2i}$ to $u_{2i}$.

The grouping information is kept by the auxiliary server, which is unknown to both the collector and individual contributors.

\textbf{Correlated perturbation in each group.} 
Each group $G_i$ of contributors then perform correlated perturbation with the help of $R_{2i-1}$ and $R_{2i}$ received from the auxiliary server. Specifically, each contributor $u_j$ generates a random variable $C_j$ independently according to the following probability distribution
\begin{equation} \label{eq:C_j}
C_j=\begin{cases}
1.5& \text{with probability $p-\sqrt{-\rho p q}$},\\
0.5& \text{with probability $\sqrt{-\rho p q}$},\\
-0.5& \text{with probability $\sqrt{-\rho p q}$},\\
-1.5& \text{with probability $q-\sqrt{-\rho p q}$},
\end{cases}
\end{equation}
for $j=2i-1$ and $2i$, where $p,q$ and $\rho$ are given in Table~\ref{tab:jointPro_2}. 

Finally, each contributor $u_j$ determines whether to report truthfully according to the following rule
\begin{equation} \label{eq:T_j}
T_j=
\begin{cases}
1&\text{if $C_j+R_j>0,$}\\
0&\text{if $C_j+R_j< 0,$}
\end{cases}
\end{equation}
for all $j=1,\dots,n$.

\begin{theorem}\label{thm:practicalM}
Under the practical mechanism, for any two contributors in the same group, the joint probability distribution of the truthfulness of the two contributors' reports is equivalent to the one described in Table.~\ref{tab:jointPro_2}.
\end{theorem}
We provide the detailed proof in Appendix~\ref{appendix:proof:practicalM}.

This practical scheme can guarantee the data privacy of each individual contributor against  the auxiliary server, the data collector, and any other contributor in the system. First, while the auxiliary server knows the which contributor receives $-1$ or $1$ and the membership information, it has no
access to perturbed value submitted by individual contributors. Similarly, the data collector receives the perturbed
values from contributors but is unaware of the group membership. Even in the worst case where all contributors but one victim contributor say $u_j$ collude with the data collector, the data collector can only infer the random variable $R_j$ that $u_j$ received from the auxiliary server but cannot recover $T_j$ because $C_j$ is unknown to the data collector. While this can lead to reduced data privacy guarantee for $u_j$, such cases of collusion are extremely impractical in a system with a large number of contributors. Last but not the least, each contributor knows only whether the other group member receives $-1$ or $1$ from the auxiliary server and is unaware of the identity of or
the perturbed value submitted by the other group member.

{\subsubsection{An MPC-Based Instantiation}
Now we introduce another practical instantiation that utilizes secure multi-party shuffling (SMPS) \cite{DBLP:journals/iacr/MovahediSZ15} to eliminate the need for a non-colluding server. SMPS is a secure multiparty computation protocol that allows a group of contributors to agree on a random permutation of their individual inputs while keeping the inputs private. After permutation, each contributor receives only one of the shuffled inputs, and no one can determine the complete mapping between the original and shuffled inputs.  Our second instantiation uses SMPS to securely shuffle contributors IDs (i.e., original positions) at random and achieving random grouping and group membership private according to the shuffled IDs. The detailed steps are as follows.}

\begin{enumerate}
    \item Upon joining the system, each contributor is assigned a unique ID $j\in\{1,2,\dots,n\}$ from the data collector. 
    \item Assuming an even number $n$ of participating contributors in a specific round, all contributors perform SMPS (Algorithm 1 in [34]) to securely shuffle their IDs, i.e., a random permutation $\pi:\{1,2,\dots,n\}\rightarrow \{1,2,\dots,n\}$. After the permutation, each contributor $j$ gets a shuffled ID $\pi(j)$ without learning any other contributor's shuffled ID.
    \item Let $\pi^{-1}(\cdot)$ be the inverse permutation of $\pi(\cdot)$.
    Every two contributors with adjacent shuffled IDs form a group without knowing each other's original ID, i.e., the $k$th group consists of contributors $\pi^{-1}(2k-1)$ and $\pi^{-1}(2k)$ for each $k\in\{1,\dots,n/2\}$.
    \item Each contributor with shuffle ID $\pi^{-1}(j)$ sets $R_j=1$ if $\pi^{-1}(j)$ is even and $-1$ if $\pi^{-1}(j)$ is odd. 
    \item Each contributor with shuffle ID $\pi^{-1}(j)$ generates $C_j$ according to Eq.~(\ref{eq:C_j}), computes $T_j$ according to Eq.~(\ref{eq:T_j}). Finally, each contributor randomly perturbs their data according to $T_j$ for submission.
\end{enumerate}

This scheme ensures random grouping due to the randomness introduced by SMPS. Moreover, Step 3 guarantees that in every group $k$, consisting of data contributors $\pi^{-1}(2k-1)$ and $\pi^{-1}(2k)$, one contributor has 
$R_{2k-1}=1$ while the other has $R_{2k}=-1$, effectively replicating the role of the trusted auxiliary server in the first scheme. Additionally, this approach ensures individual data privacy  against both the data collector and other participants. Since each contributor $j$ only knows his shuffled position $\pi(j)$ without knowing the other group member, group membership privacy is maintained. This setup allows each contributor to randomly decide whether to report truthfully, enhancing privacy while maintaining the functionality of the first scheme. On the other hand, this scheme requires all contributors to be online and participate in SMPS, leading to higher computation and communication overheads. In particular, SMPS has a computation and communication
overheads of $\mathcal{O}(n\log n)$, 
which is much higher than the first scheme. 

It is also worth noting that SMPS and data submission can be conducted at different times. We anticipate that the JRR scheme will be implemented as a mobile app, while SMPS will run as a background service, executing periodically without contributor involvement. Upon successfully participating in SMPS, contributors can submit their data via JRR at any time as needed. Newly registered contributors who have not yet completed any SMPS procedure can still submit data using classical RR. Similarly, if a contributor's group peer fails to submit their value, it effectively reduces to that contributor submitting his value under RR.

\subsection{Discussions}

In this subsection, we examine several key issues related to JRR and its practical instantiations.
\subsubsection{Extension to Non-Binary Data} \label{sec:extension-non-binary}
It is possible to extend JRR to support non-binary data by redesigning the joint reporting probability in Table~\ref{tab:jointPro_2}. In particular, for a group of two contributors with data $v_1, v_2\in [k]$, where $k\geq2$ is the domain size, they report their data according to the following joint probability distribution. 
\begin{equation}\label{eq:k-JRR}
    \Pr[(v_1', v_2')] = \begin{cases}
        p^2 + \rho pq & \text{if } v_1' = v_1, v_2' = v_2, \\
        pq - \frac{1}{k-1}\rho pq & \text{if } v_1' = v_1, v_2' \neq v_2, \\
        pq - \frac{1}{k-1}\rho pq & \text{if } v_1' \neq v_1, v_2' = v_2, \\
        q^2 + \frac{1}{(k-1)^2}\rho pq & \text{if } v_1' \neq v_1, v_2' \neq v_2,
    \end{cases}
\end{equation}
where $p + (k-1)q = 1$ and $\rho$ is the correlation coefficient between $T_1$ and $T_2$.
This is similar to the extension from RR to Generalized RR (GRR). 

It is easy prove that the marginal probability distribution of each contributor reporting his value is the same as the binary case,  i.e. $k$-$\mathrm{JRR}$ maps each value $v$ to itself with probability $p$ and to any other value with probability $q$.  As a result, $\hat{n}_v = (I_v - nq)/(p - q)$ is also an unbiased estimator for $n_v$, where $I_v$ is the number of contributors reporting $v$.
We prove this property in Appendix~\ref{appendix:proof:extension}. 

The data privacy and utility can be analyzed in the same way as the binary case.
The remaining task is to choose $p$ and $\rho$ by solving an optimization problem similar to the one in Eq.~(\ref{eq:opt_pro}), which we leave as our future work.

\subsubsection{Integration with Advanced LDP Mechanisms}

JRR can be integrated with advanced LDP mechanisms built upon RR to improve their utility and privacy tradeoff. Here we present its integration with Optimized Unary Encoding (OUE) \cite{WangLoc17} and Optimized Local Hashing (OLH) \cite{WangLoc17} as two examples.

\textbf{Integration with OUE:}
OUE is an LDP mechanism that encodes a data value in a $k$-size domain into $k$-bit binary vector \cite{WangLoc17}.
OUE follows a three-step encoding-perturbation-aggregation procedure:
(i) Encoding: given an original value $x\in[k]$, OUE first encodes it into $\mathsf{Encode}(x) =\{0, \cdots, 0, 1, 0, \cdots, 0\}$,
where the $x$-th bit is $1$ and the rest are $0$.
(ii) Perturbation: given the $k$-bit vector $B=\mathsf{Encode}(x)$, OUE then generates a perturbed vector $B'$ using RR according to the following probability distribution 
\begin{equation*}
    \Pr[B'[j] = 1] = \begin{cases}
        p, & \text{if } B[j] = 1, \\
        q, & \text{if } B[j] = 0,
    \end{cases}
\end{equation*}
where $B[j]$ denotes the $j$-th bit of $B$.
(iii) Aggregation: the data collector counts the number of $1$ in $B'$ from all data contributors whereby to estimate the frequency of each value as in RR.

JRR can be seamlessly integrated into OUE by modifying the perturbation step. Specifically, instead of applying independent RR, two contributors jointly perturb each bit of their encoded vectors using JRR. This collaborative approach enhances the estimation accuracy of OUE while preserving the original privacy guarantees.

\textbf{Integration with OLH:}
OLH follows a similar encoding-perturb\-ation-aggregation procedure: (i) Each contributor randomly picks one hash function $H(\cdot)$ among a universal hash family to map his value $v$ in an $s$-size domain into a much smaller domain of size $k$, i.e., $x=H(v)$. (ii) Each contributor then perturbs the encoded value $x$ into $y$ using GRR according to the following probability distribution
\begin{equation*}
     \Pr[y = i] = \begin{cases}
        p, & \text{if } i=x, \\
        q, & \text{for each } i\in\{1,\dots,k\}\setminus\{x\}.
    \end{cases}
\end{equation*}
(iii) 
The data collector then counts, for each 
$i\in[k]$, the number of reports equal to 
$i$, to estimate the frequency of $i$. This frequency also serves as an estimated frequency of original values that are mapped to $i$ (i.e., for which $i=H(v)$).

We can easily integrate JRR with OLH by replacing the perturbation procedure in Step~(ii) with the extended JRR for non-binary data introduced in the previous subsection, i.e., $k$-JRR given in Eq.~\ref{eq:k-JRR}, to enhance the estimation accuracy of OLH while preserving its original privacy guarantees.

\subsubsection{Impact of Large Domain Size}
Like other existing LDP mechanisms, the data utility of JRR inherently declines as the data domain size increases. This is because any LDP mechanism must allocate probability mass across all possible data values to satisfy $\epsilon$-LDP, reducing the probability that each data contributor reports their true value as the domain expands. While the extended JRR method introduced in Section~\ref{sec:extension-non-binary} i.e., $k$-JRR, consistently outperforms GRR (a special case of $k$-JRR with $\rho = 0$) by achieving a better utility-privacy trade-off through tuning the parameter $\rho$, its advantage over GRR diminishes as the domain size grows.  

However, since JRR is designed to replace the RR or GRR component in other LDP mechanisms—many of which incorporate domain reduction techniques to mitigate the impact of large domain sizes—JRR can effectively handle large domains as long as the underlying LDP mechanisms can before being enhanced by JRR. This capability is demonstrated in the integration of JRR with OLH discussed earlier.

\subsubsection{Extension to Larger-size Group}
While this paper focuses on two-contributor groups, JRR can theoretically be extended to larger groups with more than two contributors. For a group of $k>2$ contributors, their joint probability distribution of truthful reporting can be represented by a $k$-dimensional table. 
Intuitively, increasing the group size could enhance JRR's privacy-utility trade-off compared to two-contributor groups by allowing for greater correlation, which can help mitigate the added noise. As the group size continues to increase, the probability of a group including colluding contributors grows significantly, leading to additional privacy leakage and limiting further improvements in the utility-privacy trade-off.

However, designing such a scheme becomes increasingly complex as the group size grows. Consider a group of three contributors as an example. To ensure that the data collector cannot infer additional information beyond standard RR by analyzing each contributor's reported value in isolation, the marginal probability of truthful reporting for each contributor must remain consistent with RR and JRR. However, fully defining their joint probability distribution requires specifying three pairwise correlation coefficients $(\rho_{12},\rho_{13},\rho_{23})$ and a triple correlation coefficient $(\rho_{123})$ to capture higher-order dependencies. As the group size increases, the number of required correlation coefficients grows exponentially. Even if we leverage symmetry to reduce the number of independent correlation parameters to one less than the group size, selecting appropriate values while maintaining privacy guarantees remains a challenging problem. Therefore, we leave the extension of JRR to larger group sizes as future work.

\subsubsection{Relationship with the Shuffle Model}

We would like to clarify the relationship between JRR and the Shuffle Model \cite{BittaPro17,ErlinAmp19,BallePri19,CheuDis19,DBLP:conf/ccs/LuoW022,DBLP:journals/iacr/GhaziGKPV19} which
also uses a trusted server to improve the utility-privacy trade-off. In the shuffle model, data contributors perturb their data using an LDP mechanism and send the perturbed data to a trusted shuffler, which shuffles all the received data before forwarding them to the data collector. It has been shown that randomly shuffling the data can improve data privacy without sacrificing any data utility. 

We stress that JRR is not intended to replace the shuffle model. 
Instead, they can be easily integrated to further improve data privacy. Specifically, each contributor can first perturb their data via JRR and then send them to a shuffler, which in turn shuffles all the received data values before forwarding them to the data collector. The data collector can estimate $n_v$ using the same estimator as JRR. 

\section{Performance Evaluation} \label{Sec:Eval}
 
This section thoroughly evaluates the performance
of the proposed JRR mechanism using both real and synthetic datasets.

\subsection{Datasets and Simulation Setting}

We use four real-world datasets, Kosarak~\cite{Kosarak}, Amazon Rating~\cite{Amazon}, E-commerce~\cite{E-Commerce}, Census~\cite{census}, for performance evaluation.
Detailed descriptions of them are shown in Appendix~\ref{appendix:datasets}.
In addition to these four real datasets, we also generate
synthetic datasets with $n$ varying from $20$ to $2\times 10^6$ and $n_1/n$ varying from $0$ to $1$. Table~\ref{tab:dataset} summarizes these datasets.

\begin{table}[t]
	\small
	\centering
	\caption{Summary of Datasets.} \label{tab:dataset}
	\begin{tabular}{ lrrr }
	 \toprule
	 Dataset & \makecell[r]{Total\\ ($n$)} & \makecell[r]{\# of ``$1$''\\ ($n_1$)} & \makecell[r]{Pct. of ``$1$''\\($n_1/n$)} \\
	 \midrule
	 Kosarak & $2\times 10^4$ & $659$ & $0.033$\\
	Amazon & $1\times 10^4$ & $762$ & $0.076$ \\
	E-commerce & $23,486$ & $19,314$ & $0.822$\\
	Census & $1\times 10^4$ & $9,528$ & $0.953$\\
	\midrule
	Synthetic & $20\sim 2\times 10^6$ & $0 \sim 2\times 10^6$ & $0\sim 1.0$\\
	\bottomrule
	\end{tabular}
\end{table}

\begin{table}[t]
		\small
			\centering
			\caption{Default Simulation Setting} \label{tab:default}
			\begin{tabular}{ lrr }
				\toprule
				Parameter & Value & Description                 \\
				\midrule
				$n$                & $10,000$       & \# of participated contributors     \\
		
				$n_1/n$                & $0.1$          & Ratio of contributors with value $1$ \\
		
				$\varepsilon$         & $0.1$          & Privacy budget                       \\
				$M$                & $5 $           & \# of colluding contributors                \\
		
				$\triangle p$      & $0.0001$      & Search granularity                   \\
		
				$\triangle{\rho}$  & $0.0001$      & Search granularity                   \\
				\bottomrule
			\end{tabular}
\end{table}

We compare the proposed JRR mechanism with the RR mechanism because RR is not only the most classical LDP protocol for frequency estimation but also a special case of JRR. 
We do not compare JRR with the shuffle model because a fair comparison between them is challenging for two reasons. First, the privacy guarantee offered by the shuffle model is derived based on the $(\varepsilon, \delta)$-DP definition with $\delta \neq 0$ ~\cite{CheuDis19,DBLP:conf/ccs/LuoW022,BallePri19,ErlinAmp19}, whereas JRR provides $\varepsilon$-LDP, i.e., $\delta=0$. It is therefore difficult to compare their estimation accuracy under the same data privacy guarantee. Second, the shuffle model measures the lower bound of estimation error using the $(\alpha,\beta)$-accuracy notion \cite{DBLP:journals/iacr/GhaziGKPV19}, which is weaker than the standard mean square error we use to measure the estimation error of JRR.
Additionally, we do not compare JRR with other more advanced LDP mechanisms, as they rely on RR as a building block and target different data types, such as OLH \cite{WangAns19} for categorical data and PCKV \cite{GuPCK20} for key-value data. 
As discussed in Section~4.6.2, JRR has the potential to replace RR in these schemes and improve their privacy-utility trade-off.

Data utility comparisons are performed at the same privacy level $\varepsilon$.
For RR, utility is maximized by setting $p = e^{\varepsilon} / (1+e^{\varepsilon})$. 
For JRR, we employ the heuristic solution of $p$ and $\rho$ outlined in Algorithm~\ref{algo:search_rho}.
Notably, to ensure a fair comparison at the same privacy level $\varepsilon$, the parameter $p$ in JRR differs from that in RR.
We use the following two metrics to evaluate the performance of JRR.
\begin{itemize}
\item \emph{Mean squared error (MSE)} \cite{GuPro20,WangLoc17}: it is the mean squared errors of the estimated $\hat{n}_v$ with respect to the real one $n_v$ across all values, which is defined as
\begin{equation}
MSE=\frac{1}{|D|}\sum_{v\in D}(\hat{n}_v-n_v)^2.
\end{equation}

\item \emph{Averaged relative error (ARE)} \cite{YePri19,LiPri12}: it is the mean relative error across all values that is defined as
\begin{equation}
ARE=\frac{1}{|D|}\sum_{v\in D}\frac{|\hat{n}_v-n_v|}{n_v}.
\end{equation}
   \end{itemize}
In the above formula, $|D|=2$ is the size of $D=\{0,1\}$. 
Note that as shown in Eq.~(\ref{eq:var_rr}) the MSE of RR is not affected by $n_v$, whereas its ARE is. Consequently, RR's ARE performance may exhibit multiple distinct lines for different $n_v$, whereas its MSE remains a single line.
	
Table~\ref{tab:default} lists the default simulation settings. Using MATLAB, each point in the figures represents the average of 1000 runs with unique random seeds.

\begin{figure*}[t]
	\centering
	\subfigure[$\varepsilon=0.01$]{\label{fig:exp:reald_mse:a}
		\includegraphics[width=0.265\textwidth]{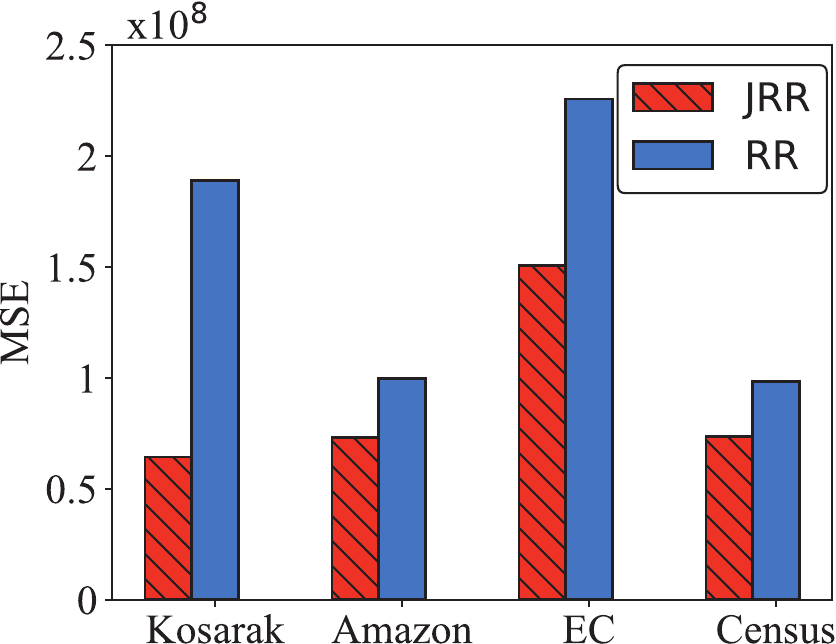}
	}\hfill
	\subfigure[$\varepsilon=0.1$] {\label{fig:exp:reald_mse:b}
		\includegraphics[width=0.265\textwidth]{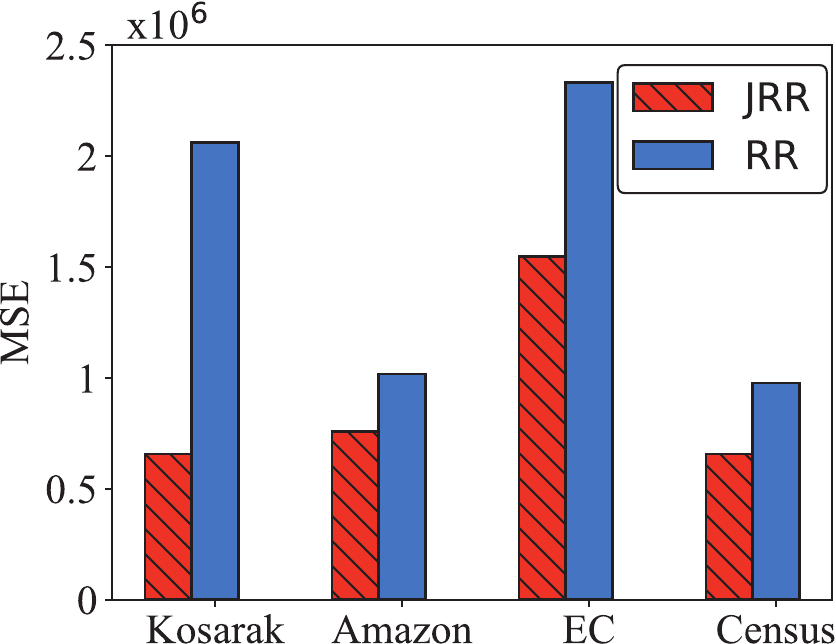}
	}
        \hfill
	\subfigure[$\varepsilon=1$]{\label{fig:exp:reald_mse:c}
		\includegraphics[width=0.265\textwidth]{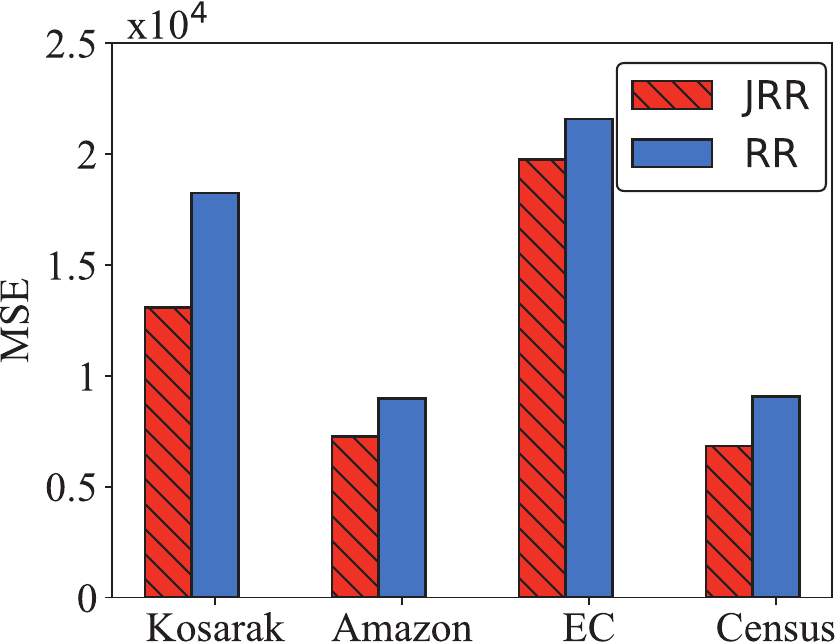}
	}
	\caption{Comparison of MSE under RR and JRR on four real datasets when the privacy budget $\varepsilon=0.01$, $0.1$ and $1$.}
	\label{fig:exp:reald_mse}
\end{figure*}

\begin{figure*}[t]
	\centering
	\subfigure[$\varepsilon=0.01$]{\label{fig:exp:reald_mre:d}
		\includegraphics[width=0.27\textwidth]{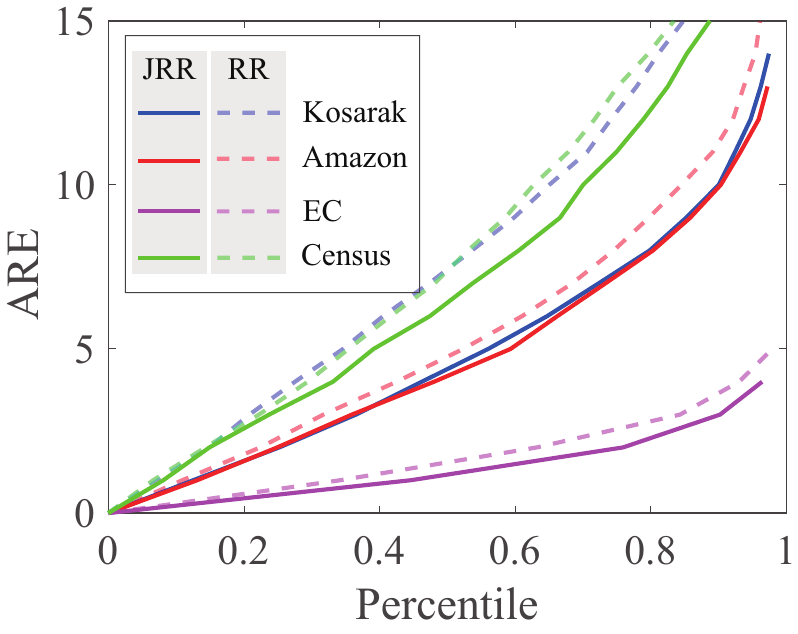}
	}\hfill
	\subfigure[$\varepsilon=0.1$]{\label{fig:exp:reald_mre:e}
		\includegraphics[width=0.27\textwidth]{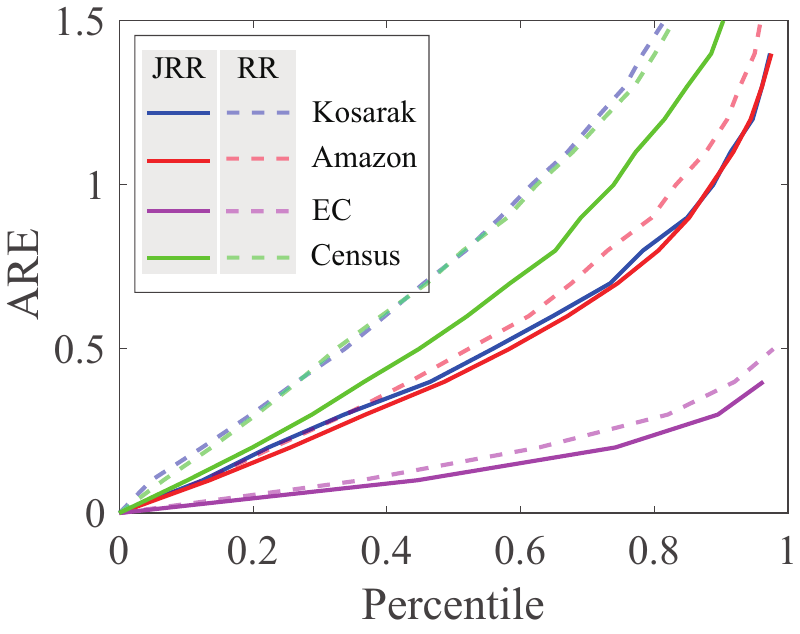}
	}
        \hfill
	\subfigure[$\varepsilon=1$]{\label{fig:exp:reald_mre:f}
		\includegraphics[width=0.27\textwidth]{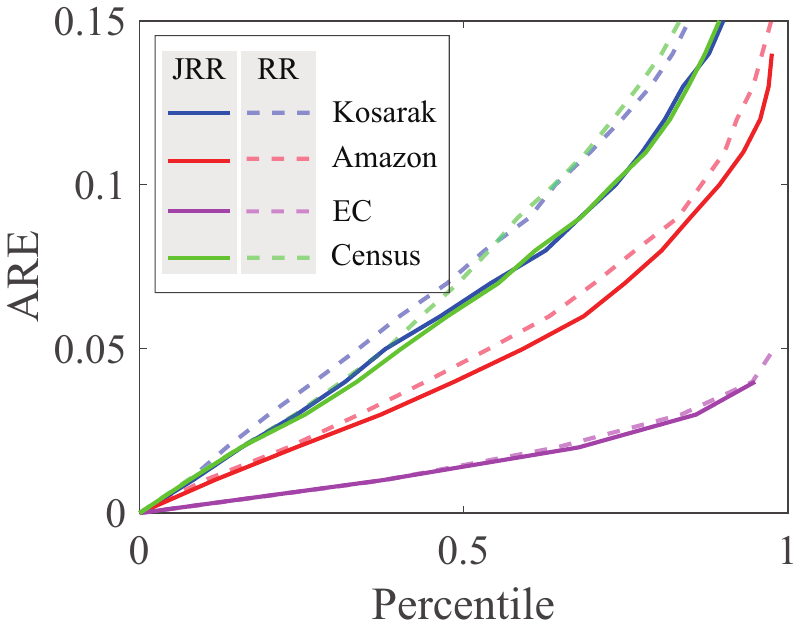}
	}
	\caption{Percentiles of ARE under RR and JRR on four real datasets when the privacy budget $\varepsilon=0.01$, $0.1$ and $1$.
	}
	\label{fig:exp:percentile}
\end{figure*}

\subsection{Results From Real-world Datasets}

Fig.~\ref{fig:exp:reald_mse} presents the MSE of JRR and RR on the four real-world datasets when $\varepsilon$ is $0.01$, $0.1$, and $1$, respectively. We can see that the JRR achieves a lower MSE than RR for all four datasets under all three $\varepsilon$s. This is expected because the negative correlation between two contributors' random perturbations under JRR can effectively reduce the expected MSE when the ratio $n_1/n$ is not close to $0.5$, which is true for all four real datasets with $n_1/n$ being either smaller than $0.1$ or larger than $0.8$. 
Moreover, we can see that the larger the $\varepsilon$, the smaller the MSE under both JRR and RR. This is also anticipated as the larger the privacy budget $\varepsilon$, the more likely that each contributor reports truthfully, the smaller the MSE under both mechanisms and vice versa. 
In addition, we can see that JRR outperforms RR by a larger margin on Kosarak and Census datasets in comparison with 
the Amazon Rating and E-commerce datasets, especially when $\varepsilon$ is small, e.g., $\varepsilon=0.01$. 
This is because the ratio, $n_1/n$, in the Kosarak and Census datasets are farther away from $0.5$ than those in the other two datasets. We will carefully evaluate the impact of $\varepsilon$, $n$, and $n_1/n$ on the MSE using the synthetic datasets shortly.

Figs.~\ref{fig:exp:percentile} show the distributions of ARE over the $1,000$ runs under JRR and RR on the four real datasets
with $\varepsilon=0.01$ and $1$, respectively.
A percentile indicates the percentage of error values that are lower than the corresponding ARE. 
We can see that for any specific percentile, JRR consistently outperforms RR with a lower ARE across all four datasets. 
For example, as shown in Fig.~\ref{fig:exp:reald_mre:e}, when $\varepsilon=0.1$, the $80$th percentile under JRR on the Kosarak dataset is $0.7$, i.e., $800$ out of $1,000$ runs have ARE lower than $0.7$. 
In contrast, the $80$th percentile under RR is $1.45$. 
These results confirm that JRR consistently offers stable performance with a lower ARE.

\subsection{Results From Synthetic Datasets}\label{subsec:syn_data}

\begin{figure*}[th!]
	\centering
	\subfigure[$n=1\times 10^4$]{\label{fig:exp:mse_epsilon_n_10k}
		\includegraphics[width=0.27\textwidth]{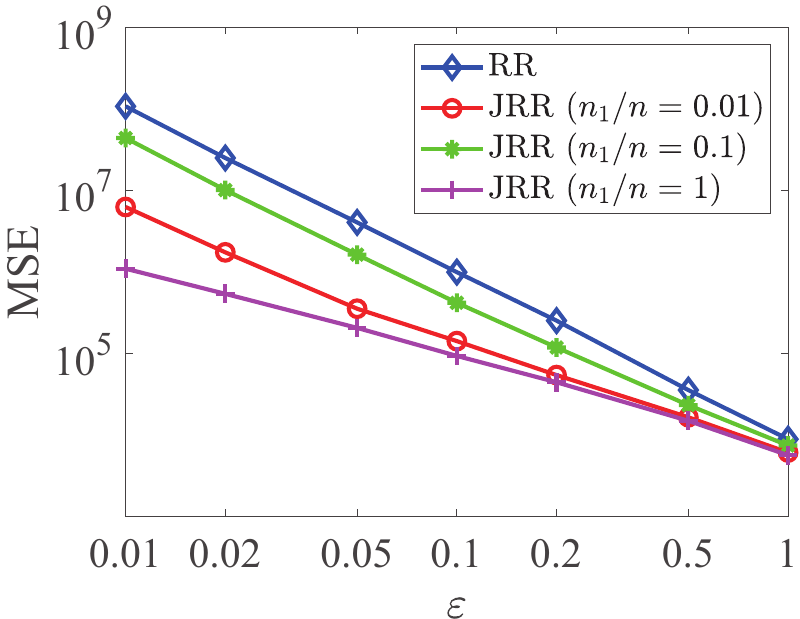}
	}\hfill
	\subfigure[$n=4\times 10^4$]{\label{fig:exp:mse_epsilon_n_40k}
		\includegraphics[width=0.27\textwidth]{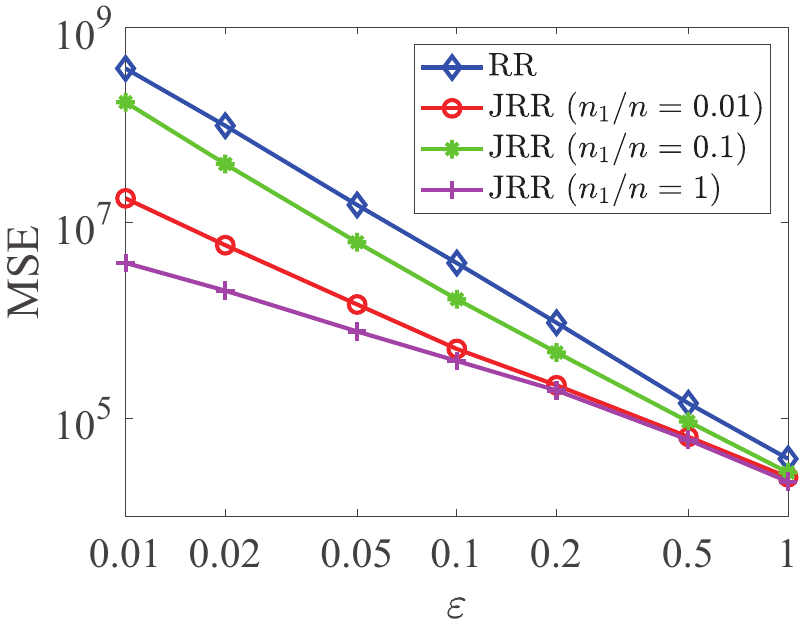}
	}\hfill
	\subfigure[$n=8\times 10^4$]{\label{fig:exp:mse_epsilon_n_80k}
		\includegraphics[width=0.27\textwidth]{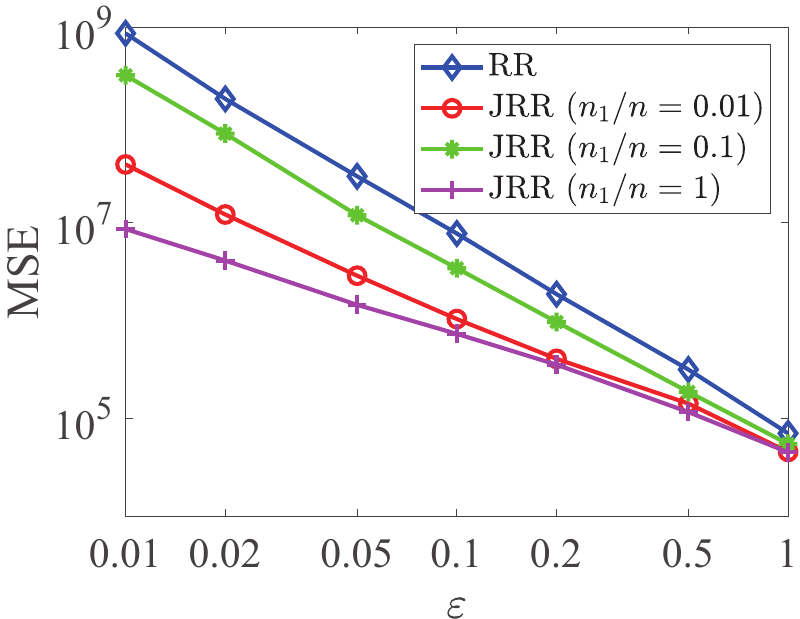}
	}
	\subfigure[$n=1\times 10^4$]{\label{fig:ARE_e_n10k}
		\includegraphics[width=0.27\textwidth]{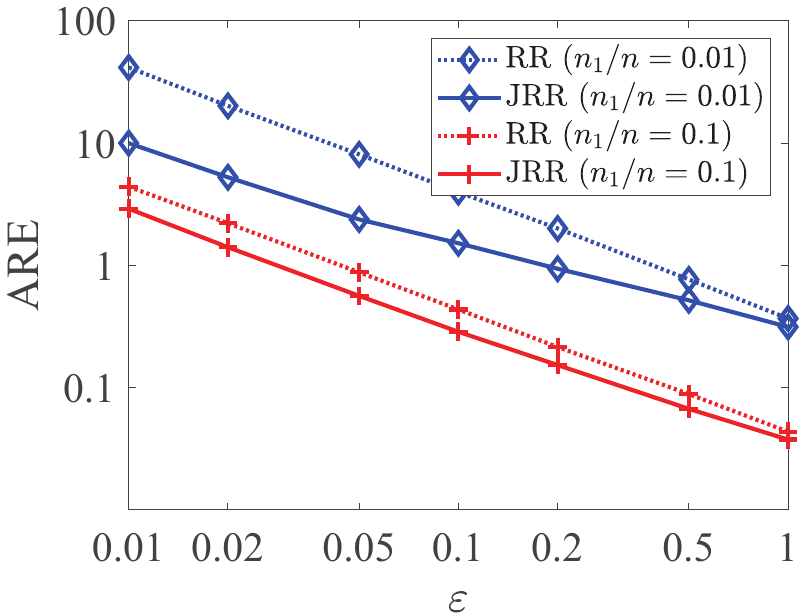}
	}\hfill
	\subfigure[$n=4\times 10^4$] {
		\includegraphics[width=0.27\textwidth]{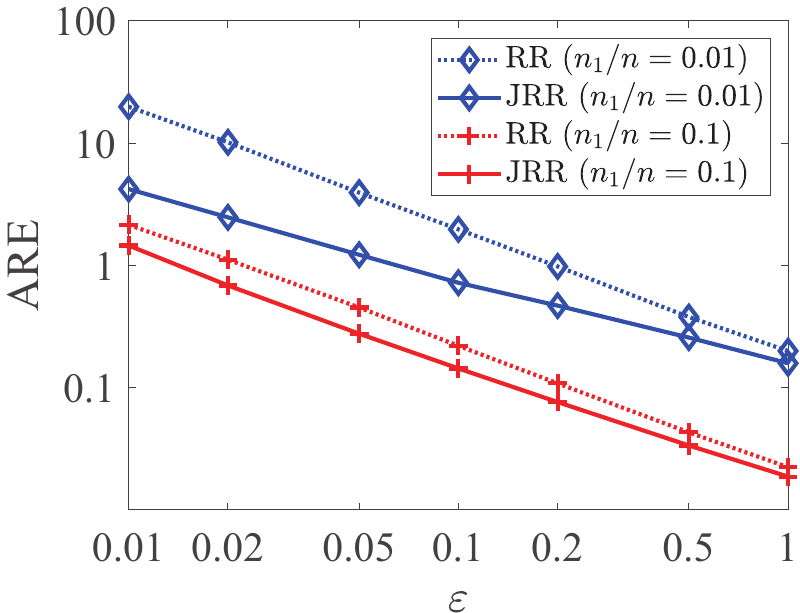}
	}\hfill
	\subfigure[$n=8\times 10^4$]{\label{fig:ARE_e_n80k}
		\includegraphics[width=0.27\textwidth]{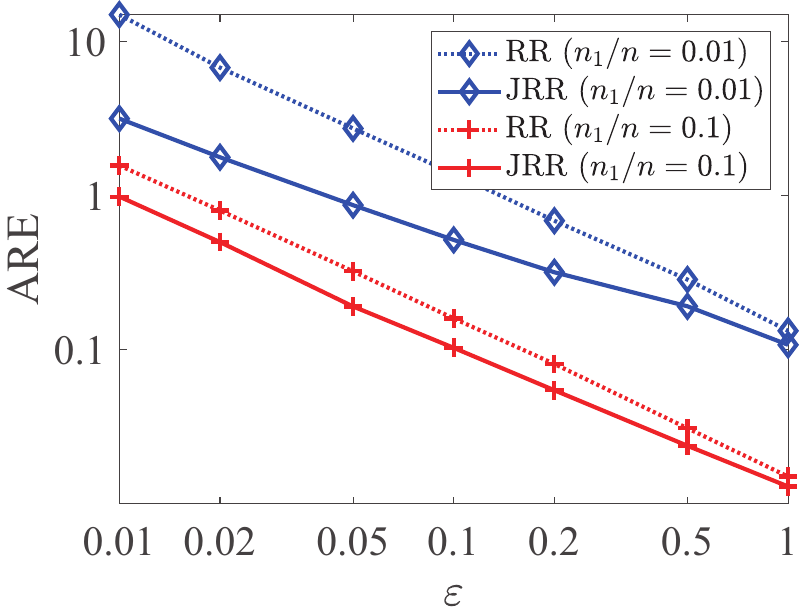}
	}
	\caption{Comparison of MSE (top row) and ARE (bottom row) under RR and JRR with privacy budget {$\varepsilon=0.01$ to $1$.}} 
	\label{fig:exp:mse_epsilon}
\end{figure*}

\begin{figure*}[th!]
	\centering
	\subfigure[$\varepsilon = 0.01$]{\label{fig:exp:mse_n:a}
		\includegraphics[width=0.27\textwidth]{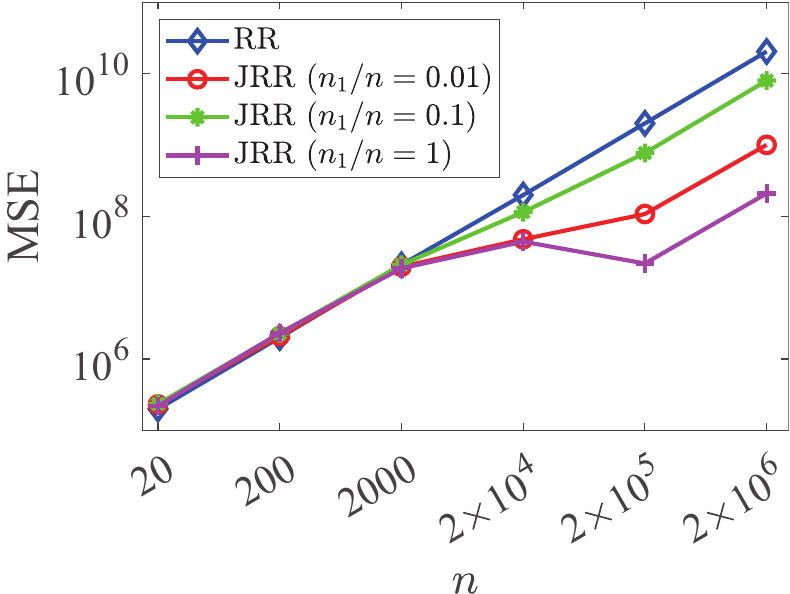}
	}\hfill
	\subfigure[$\varepsilon = 0.1$] {\label{fig:exp:mse_n:b}
		\includegraphics[width=0.27\textwidth]{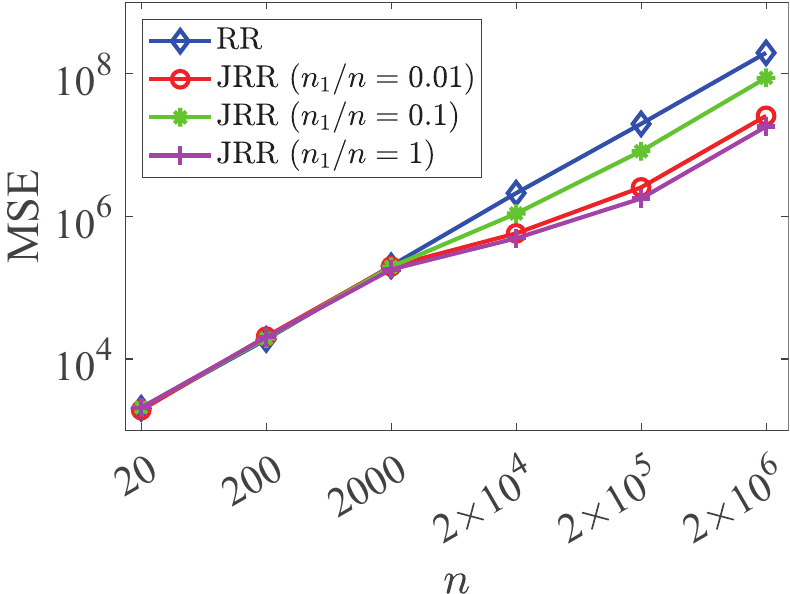}
	}\hfill
	\subfigure[$\varepsilon = 1$]{\label{fig:exp:mse_n:c}
		\includegraphics[width=0.27\textwidth]{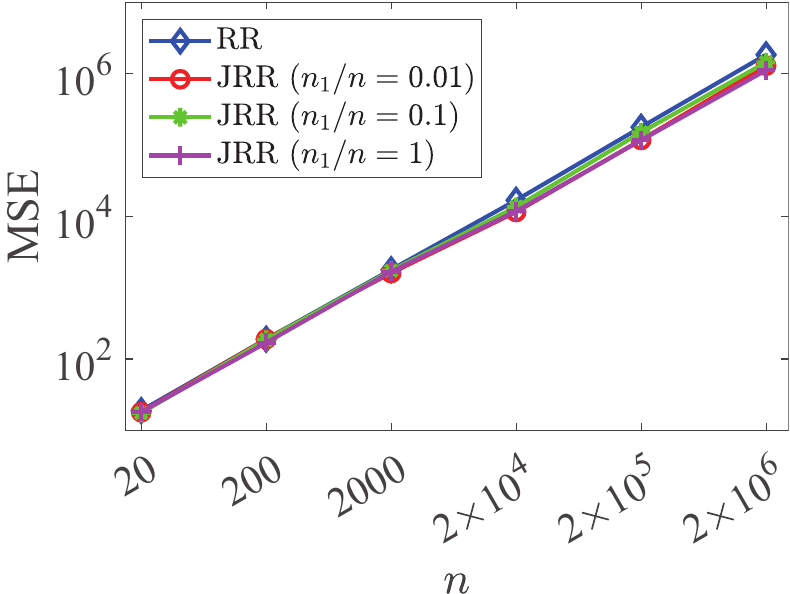}
	}
	\subfigure[$\varepsilon = 0.01$]{\label{fig:exp:are_n:a}
		\includegraphics[width=0.27\textwidth]{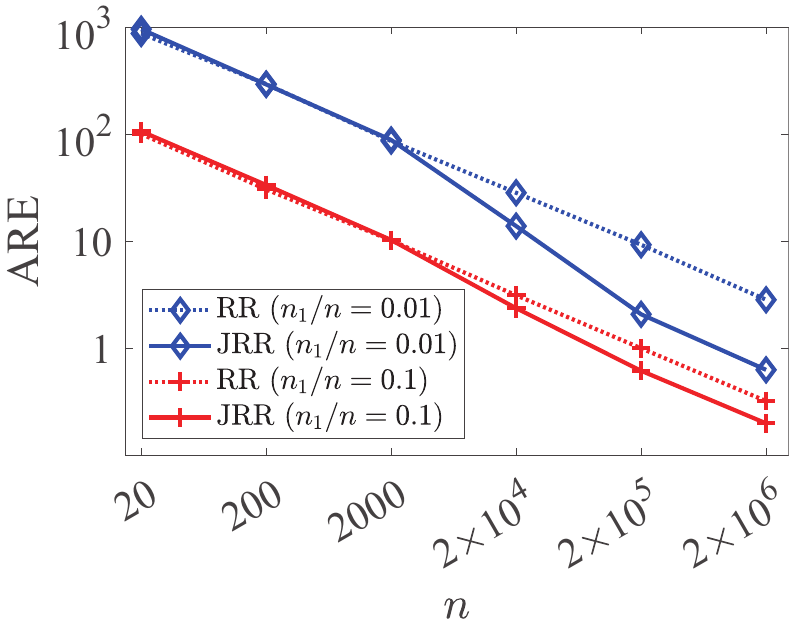}
	}\hfill
	\subfigure[$\varepsilon = 0.1$] {\label{fig:exp:are_n:b}
		\includegraphics[width=0.27\textwidth]{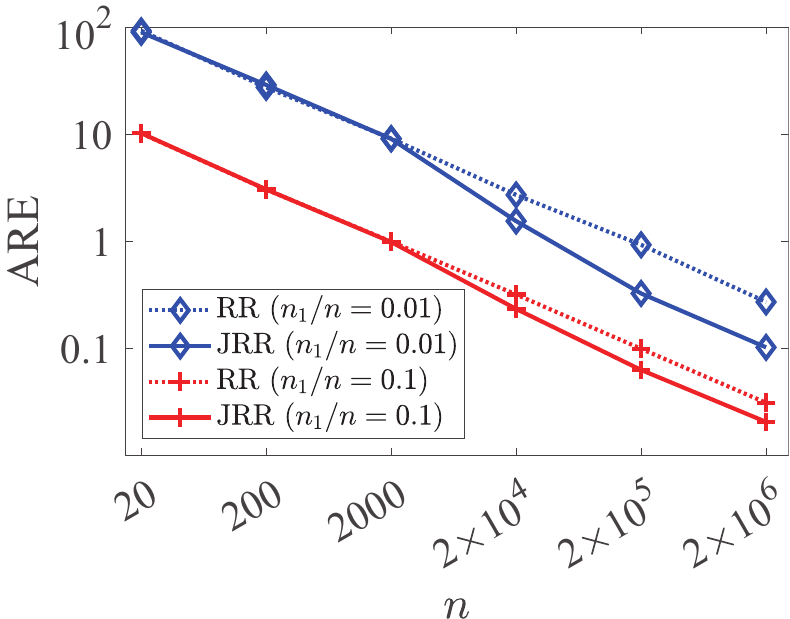}
	}\hfill
	\subfigure[$\varepsilon = 1$]{\label{fig:exp:are_n:c}
		\includegraphics[width=0.27\textwidth]{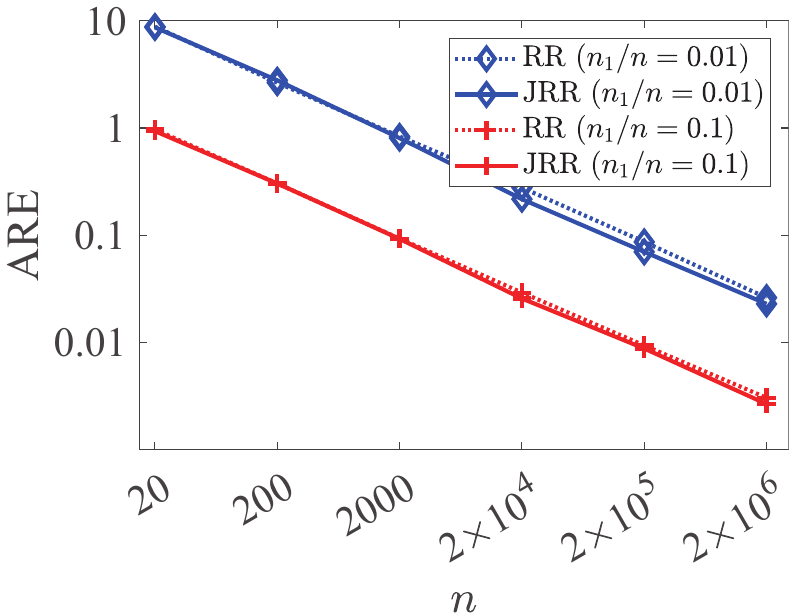}
	}
	\caption{Comparison of MSE under RR and JRR with $n=20$ to $2\times 10^6$.} 
	\label{fig:exp:mse_n}
\end{figure*}

\begin{figure*}[th!]
\centering
\hspace{0.25em}
\subfigure[$n=1\times 10^4$]{\label{fig:mse_m_a}
	\includegraphics[width=0.26\textwidth]{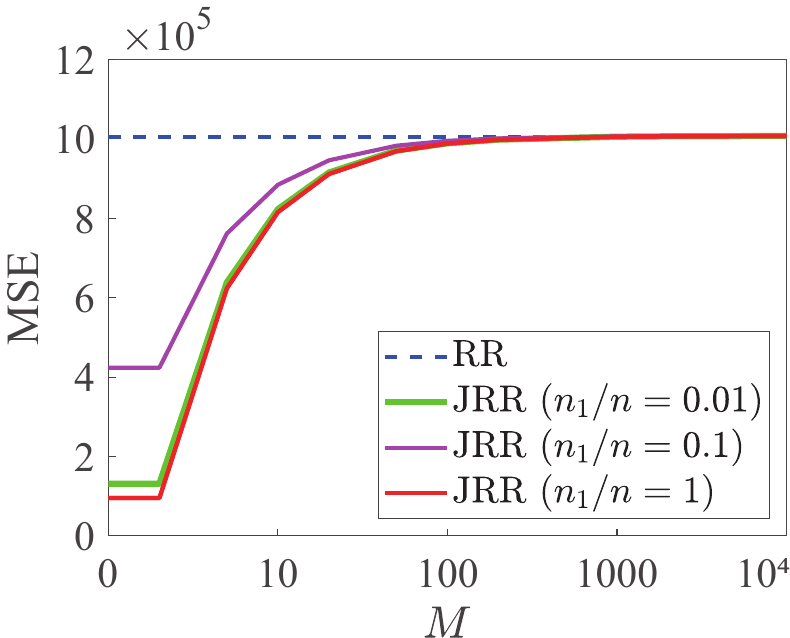}
}\hfill \hspace{0.15em}
\subfigure[$n=4\times 10^4$] {\label{fig:mse_m_b}
	\includegraphics[width=0.26\textwidth]{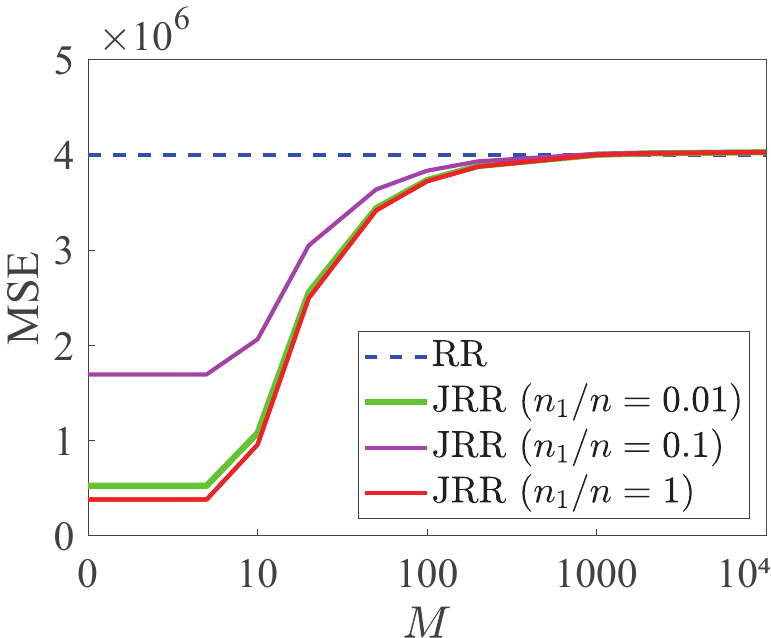}
}\hfill
\subfigure[$n=8\times 10^4$]{\label{fig:mse_m_c}
	\includegraphics[width=0.26\textwidth]{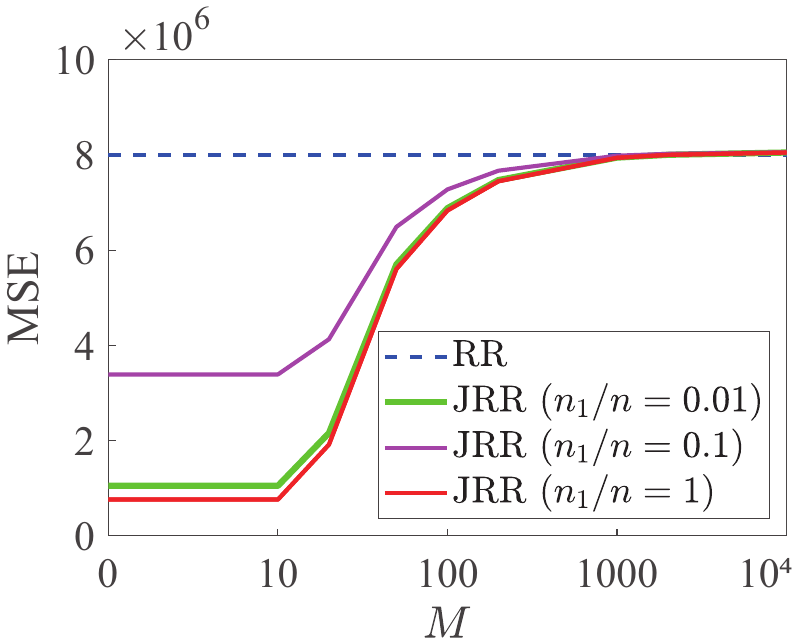}
}

\caption{Comparison of MSE of JRR's heuristic solution with RR.}
\label{fig:exp:heuristic_effect_m}
\end{figure*}

\subsubsection{Impact of $\varepsilon$}
Figs.~\ref{fig:exp:mse_epsilon_n_10k} to \ref{fig:exp:mse_epsilon_n_80k} illustrates the MSEs of RR and JRR with privacy budgets $\varepsilon$ varying from 0.01 to 1 under different number of data contributors. As expected, MSE decreases for both RR and JRR as $\varepsilon$ increases because higher $\varepsilon$ increases the probability of reporting truthfully.  Moreover, JRR consistently achieves a smaller MSE than RR, with the performance gap widening as $n$ increases.
For example, as shown in  Fig.~\ref{fig:exp:mse_epsilon_n_10k}, when $\varepsilon=0.01$, the MSE under JRR is $5.8\%$, $40.9\%$, and $1.0\%$ of that under RR when $n_1/n$ is 0.01, 0.1, and 1, respectively. In contrast, when $n = 8 \times 10^4$ (see Fig.~\ref{fig:exp:mse_epsilon_n_80k}), the corresponding MSE is $4.5\%$, $37.2\%$, and $0.9\%$ of that under RR.
This trend occurs because a larger $n$ reduces the likelihood of the data collector correctly identifying group members, limiting the additional information inferred from correlated reporting.
The advantage of JRR over RR becomes even more pronounced as $\varepsilon$ decreases.
In Fig.~\ref{fig:exp:mse_epsilon_n_80k}, for $\varepsilon=0.1$, the MSE under JRR is $86.6\%$, $55.8\%$, and $90.9\%$ smaller than RR's when $n_1/n$ is 0.01, 0.1, and 1, respectively. In contrast, when $\varepsilon=0.01$, the corresponding MSE is $95.4\%$, $62.8\%$, and $99.0\%$ smaller than that under RR, respectively. 
When $\varepsilon=0.1$ and $n_1/n=1$, the MSE under JRR is $7.3\times 10^5$, which is about $74.1\%$ of the one under RR. 
In contrast, when $\varepsilon=0.01$, the MSE under JRR $n_1/n=1$ is $8.6\times 10^6$, which is only about $1.0\%$ of the one under RR. These results demonstrate that JRR outperforms RR with a large margin, especially when $n$ is large and $\varepsilon$ is small. 

Figs.~\ref{fig:ARE_e_n10k} to \ref{fig:ARE_e_n80k} compare the AREs under RR and JRR with $\varepsilon$ varying from $0.01$ to $1$. 
We can observe similar trends to Figs.~\ref{fig:exp:mse_epsilon_n_10k} to \ref{fig:exp:mse_epsilon_n_80k} that the AREs under both RR and JRR decreases as $\varepsilon$ increases. 
Moreover, we can see that a larger $n_1/n$ leads to a smaller ARE.
JRR achieves a smaller ARE than RR in all the cases, and the margin by which JRR outperforms RR increases as $\varepsilon$ decreases due to the same reasons that we mentioned earlier.

\subsubsection{Impact of $n$}

Figs.~\ref{fig:exp:mse_n} compares the MSE and ARE under JRR and RR with the $n$ varying from 20 to 2, 000, 000 for $\varepsilon$= 0.01, 0.1, and 1, respectively.

Figs.~\ref{fig:exp:mse_n:a} to \ref{fig:exp:mse_n:c} show that the MSE under 
RR increases linearly as $n$ increases, which is expected. In contrast, the MSE under JRR initially increases linearly as $n$ increases from $20$ to $2,000$, then increases at a slower rate or even decreases as $n$ increases from $2,000$ to $2,000,000$. 
This is because when there are relatively few contributors, $\rho$ needs to be close to zero to guarantee sufficient data privacy, and JRR and RR have similar MSE. 
As $n$ increases, $\rho$ output by Algorithm~\ref{al:heu} decreases, and a smaller negative correlation is introduced between the two contributors in each group, resulting in a smaller MSE than RR. Since the MSE under JRR is the sum of the variance of all $n/2$ groups, the change in the MSE under JRR is the joint effect of the decreased variance in each group and the increased number of groups. As $n$ increases, the MSE under JRR inevitably increases but is still lower than that under RR.
Most notably, in Fig.~\ref{fig:exp:mse_n:a} when $n=200,000$ and $n_1/n=1$, JRR outperforms RR by two orders of magnitude.
Additionally, although the margin JRR outperforms RR decreases as $\varepsilon$ increases, the improvement remains significant even when $\varepsilon$ is large. Taking Fig.~\ref{fig:exp:mse_n:c} as an example, the MSE under JRR is still $7.2\times 10^5$ lower than that of RR when $n=2\times 10^6$ and $n_1/n = 1$.

\begin{figure*}[th!]
\centering
\subfigure[$n=1\times 10^4$]{\label{fig:exp:effect_n1:a}
	\includegraphics[width=0.27\textwidth]{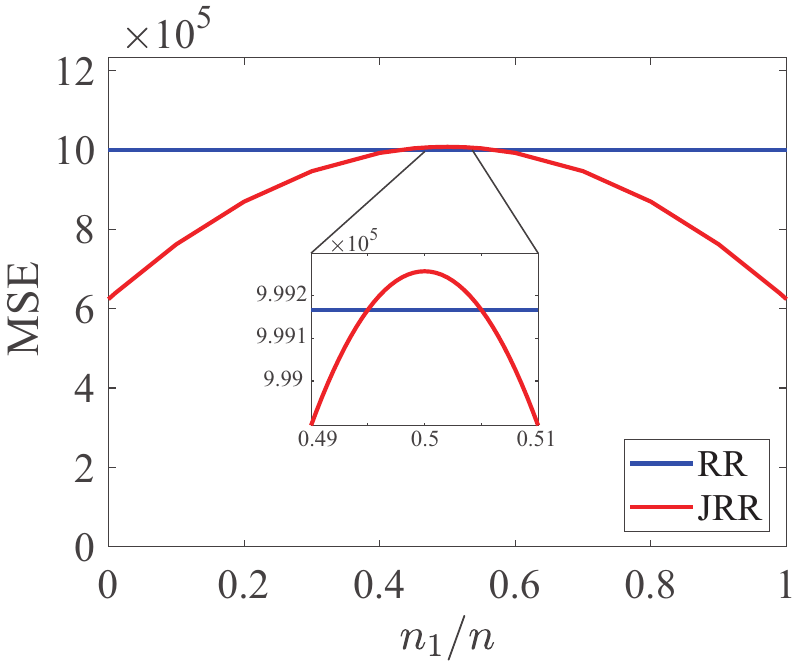}
}\hfill
\subfigure[$n=4\times 10^4$] {
	\includegraphics[width=0.27\textwidth]{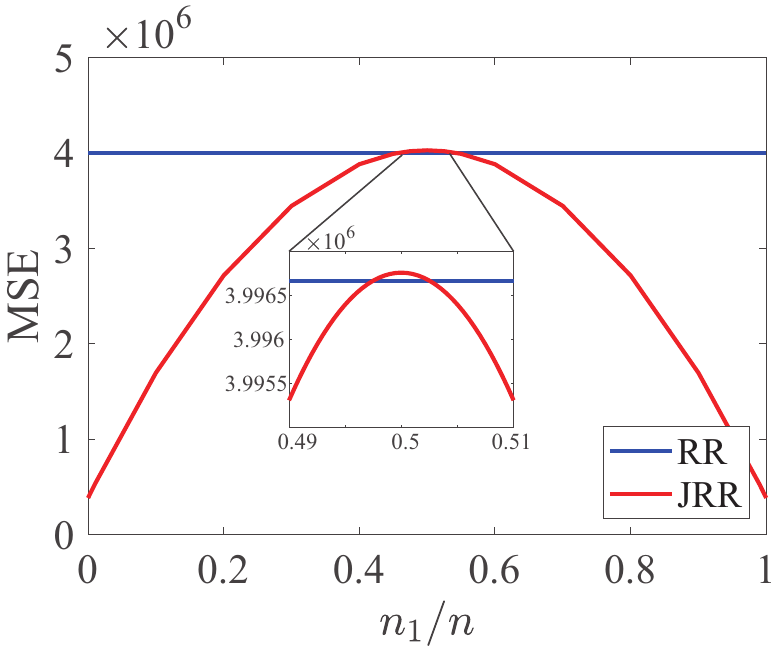}
}\hfill
\subfigure[$n=8\times 10^4$]{\label{fig:var_f_e1}
	\includegraphics[width=0.27\textwidth]{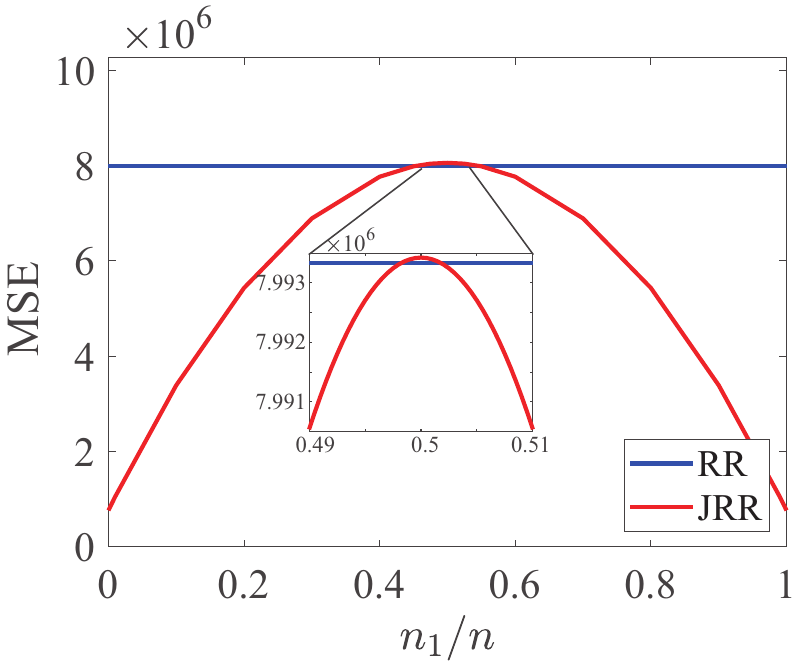}
}
\vspace{-0.2cm}
\caption{Comparison of MSE under RR and JRR with $n_1/n=0$ to $1$.}
\label{fig:exp:effect_n1_n}
\end{figure*}

\begin{figure*}[th!]
	\centering
	   \begin{minipage}{0.49\textwidth}
		  \centering
	   \subfigure[RI vs. $\varepsilon$] {\label{fig:worst_e}
		   \includegraphics[width=0.47\textwidth]{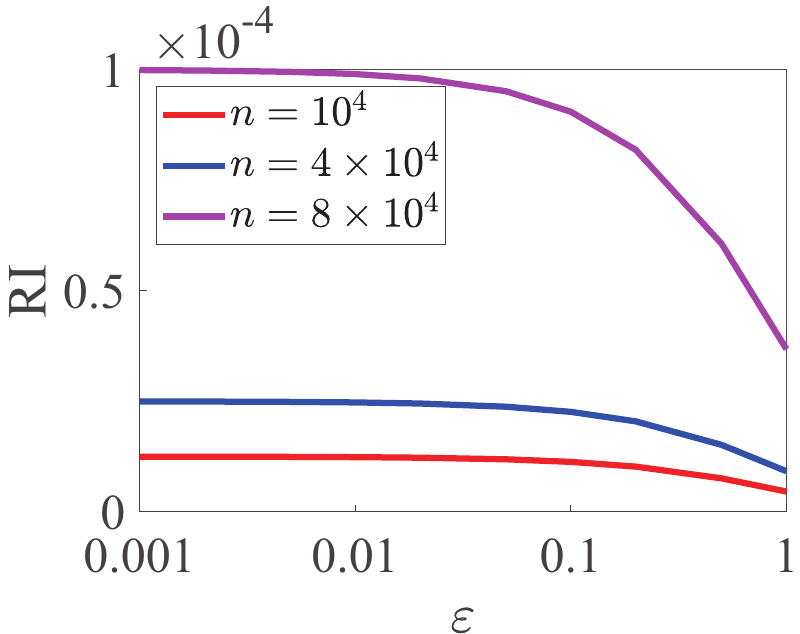}
	   }\hfill
	   \subfigure[$R$ vs. $\varepsilon$]{\label{fig:rangf_e}
		   \includegraphics[width=0.47\textwidth]{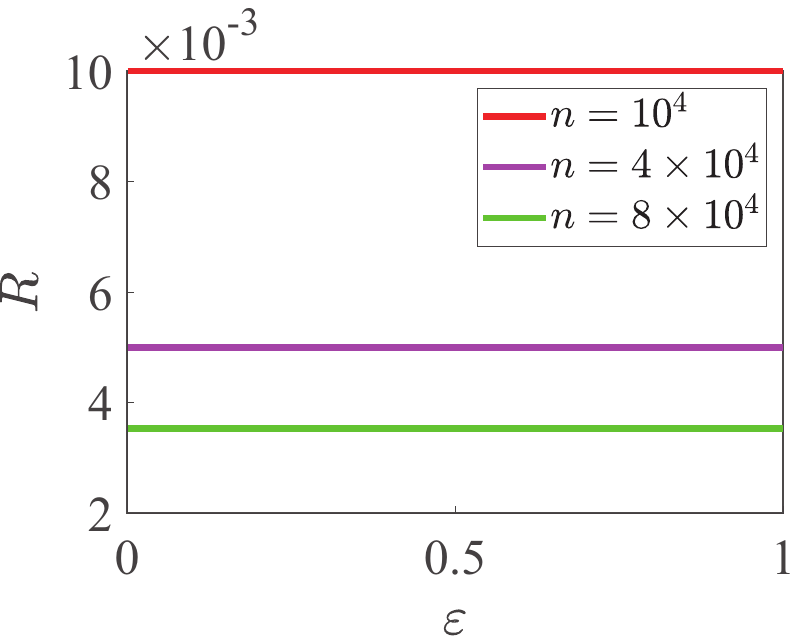}
	   }
	   \caption{The impact of $\varepsilon$ on RI and $R$.}\label{fig:e_under}
	   \end{minipage}\hfill
	   \begin{minipage}{0.49\textwidth}
			 \subfigure[RI vs. $n$] {\label{fig:worst_n}
		   \includegraphics[width=0.47\textwidth]{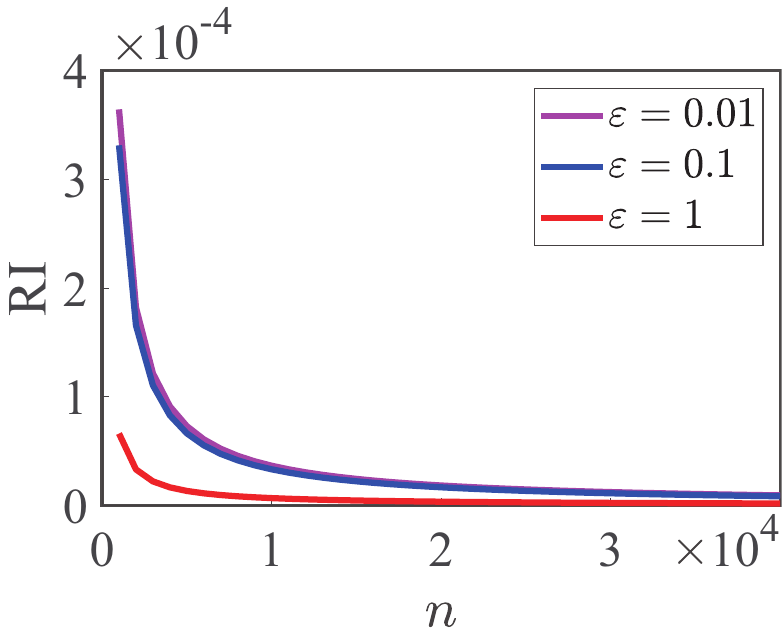}
	   }\hfill
	   \subfigure[$R$ vs. $n$]{\label{fig:rangf_n}
		   \includegraphics[width=0.47\textwidth]{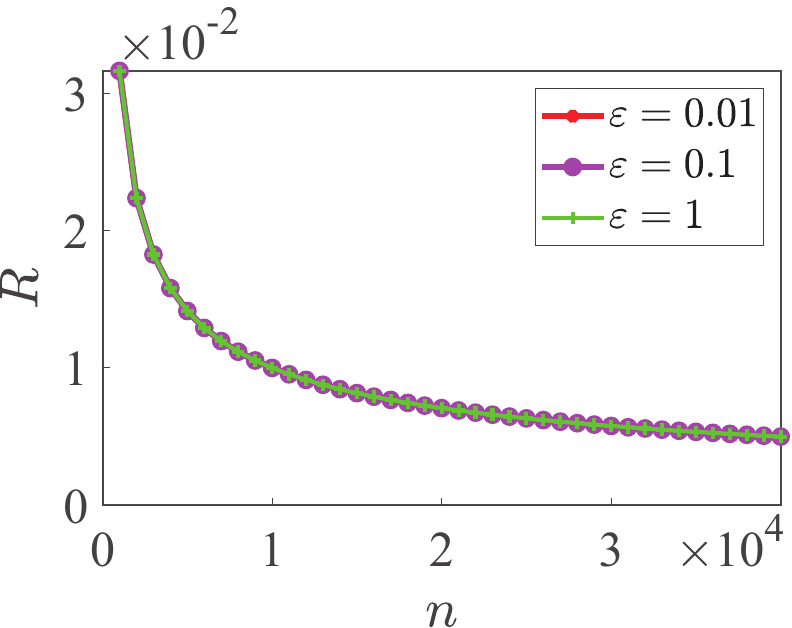}
	   }
	   \caption{The impact of $n$ on RI and $R$.} \label{fig:rangf}
	   \end{minipage}
   \end{figure*}
Figs.~\ref{fig:exp:are_n:a} to \ref{fig:exp:are_n:c} show a similar trend of ARE under RR and JRR. Specifically, the ARE under JRR and RR both decrease as $n$ increases. As $n$ increases from $2,000$ to $2,000,000$, the ARE under JRR decreases much faster than that under RR due to the joint effect of increasing $n$ and decreasing variance in each group.  

These results show that JRR is particularly favorable for the cases of small $\varepsilon$, large $n$, and $n_1/n$ close to $0$ or $1$, reducing RR's MSE by up to two orders of magnitude and ARE by over 70\%. 

\subsubsection{Impact of $M$}
Figs.~\ref{fig:mse_m_a} to \ref{fig:mse_m_c} show the MSE under JRR and RR as $M$ increases from 0 to $n-1$, where the MSE under RR is not affected by $M$ and is plotted for reference.
We can see that the MSE of JRR initially almost stays stable and then increases until it reaches that of RR.
The reason is that the $p$ selected by Algorithm~\ref{al:heu} is always close to the one achieved under RR, but the corresponding $\rho$ increases as $M$ increases. Specifically, when $M$ is small, $\rho$ is always very close to the minimal value $1-1/p$, resulting in a relatively stable MSE that is much smaller than the one under RR. As $M$ increases, a small $\rho$ no longer satisfies the privacy constraint, and an increased negative $\rho$ leads to an increased MSE. When $M$ is very large, e.g., $M=90\%n$, $\rho$ is close to 0, and JRR degrades to RR.

These results indicate that JRR consistently outperforms RR in terms of MSE for any $M$ from $0$ to $n-1$, and is particularly favorable when $M$ is small.

\subsubsection{Impact of $n_1/n$}
Fig.~\ref{fig:exp:effect_n1_n} compares the MSE of JRR and RR on synthetic datasets, with $\varepsilon=1$ and $n_1/n$ ranging from $0$ to $1$.
We can see that the MSE under JRR initially increases and then decreases as $n_1/n$ increases from 0 to 1 and is symmetric with respect to $n_1/n$. The reason for the initial increase is that the MSE under JRR has term $\rho((2n_1-n)^2-n)=\rho n^2((2n_1/n-1)^2-1/n)$, which is monotonically increasing with respect to $n_1/n\in[0,0.5]$ when $\rho<0$. In addition, the symmetry comes from the fact that $MSE=(\hat{n}_1-n_1)^2=(\hat{n}_0-n_0)^2$, so MSE does not change if every contributor's original value is flipped. 
Moreover, we can see that the MSE of JRR exceeds that under RR when $n_1/n$ is close to $0.5$. There are two reasons. First, we choose $p$ and $\rho$ by Algorithm~\ref{algo:search_rho} under the assumption that $n_1/n\notin [1/2-1/2\sqrt{n},1/2+1/2\sqrt{n}]$. When $n_1/n$ is close to $0.5$, this assumption does not hold, and the choice of $p$ and $\rho$ results in higher MSE than RR. Second, the $p$ and $\rho$ chosen by Algorithm~\ref{algo:search_rho} are not the optimal solution for the optimization problem given in Eq.~(\ref{eq:opt_pro}), which may further increase the MSE.

\subsubsection{The Cases of JRR Underperforming RR}
We further evaluate the conditions under which JRR underperforms RR using the following two metrics:
\begin{itemize}
    \item \emph{Relative increases (RI):} It is defined as the ratio of the difference between the MSE of JRR and that of RR to the MSE of RR
	when $\frac{n_1}{n}=0.5$ (i.e., the worst case for JRR):
\begin{equation}
    \text{RI}=\frac{MSE_{\text{JRR}}-MSE_{\text{RR}}}{MSE_{\text{RR}}},
\end{equation}
 where $MSE_{\text{JRR}}$ and $MSE_{\text{RR}}$ are the MSE of JRR and RR, respectively.
 
\item \emph{Ratio of underperforming range ($R$):} Since the MSE of JRR and RR are roughly symmetric to $\frac{n_1}{n}=0.5$, 
the range of $\frac{n_1}{n}$ in which the MSE of JRR exceeds that of RR is $[0.5-R/2,0.5+R/2]$ for some $R\in[0,0.5]$. 
Therefore, we define $R$ as the ratio of the underperforming range. 
\end{itemize}

Figs.~\ref{fig:worst_e} and \ref{fig:rangf_e} show RI and $R$ with $\varepsilon$ varying from $0.001$ to $1$. We can see from Fig.~\ref{fig:worst_e} that RI
decreases as $\varepsilon$, which is anticipated because a larger $\varepsilon$ means a larger $p$ under both JRR and RR. Notably, RI is always less than $10^{-4}$ even for an extremely small $\varepsilon=0.001$. From Fig.~\ref{fig:rangf_e}, we can see that $R$ remain stable as $\varepsilon$ increases, but a larger $n$ (e.g., $n=80,000$) results in a smaller $R$, coinciding with theoretical analysis in Section~\ref{subsec:CRRM_opt} that the range of underperforming is $1/\sqrt{n}$, i.e., independent of $\varepsilon$ but decreases as $n$ increases.

Figs.~\ref{fig:worst_n} and \ref{fig:rangf_n} show RI and $R$ with $n$ varying from $1,000$ to $40,000$. We can see from Fig.~\ref{fig:worst_n} that RI initially decreases sharply and then gradually decreases as $n$ increases. In particular, even when $n$ is small, e.g., $n=5,000$, RI is $7.3\times 10^{-5}$, which is negligible.
Moreover, as we can see in Fig.~\ref{fig:rangf_n}, $R$ decreases as $n$ increases, but it is not affected by $\varepsilon$, which is consistent with Fig.~\ref{fig:rangf_e}. 
In addition, even when $n=1,000$, $R$ is less than $3\%$, indicating JRR outperforms RR in terms of the MSE for more than $97\%$ of value $n_1/n$. 

These results indicate that JRR outperforms RR for an overwhelming majority of $n_1/n$ with only a negligible relative increase in the worst case.

\subsection{Summary of Simulation Results}\label{subsec:summary}

We summarize the simulation results as follows. 
\begin{itemize}
    \item JRR achieves smaller MSE and ARE than RR as long as the numbers of contributors having value 1 and 0 are not very close, i.e., $n_1/n\notin [1/2-1/2\sqrt{n},1/2+1/2\sqrt{n}]$. 
\item JRR significantly outperforms RR when the numbers of contributors with values $1$ and $0$ are not close, the total number of contributors is large, and the number of colluding contributors is small.
\item When $n_1$ and $n_0$ are not very close, the margin by which JRR outperforms RR is inversely proportional to the privacy budget $\varepsilon$ and the maximum number of colluding contributors $M$ but proportional to the number of contributors $n$. As $M$ increases, the MSE of JRR approaches that of RR. 
\item JRR underperforms RR if the numbers of contributors having value 1 and 0 are very close, i.e., when $n_1/n\in [1/2-1/2\sqrt{n},1/2+1/2\sqrt{n}]$. However, the margin by which RR outperforms JRR is very small or negligible. 

\end{itemize}

\section{Related Work} \label{Sec:Related}

Privacy-preserving frequency estimation dates back to Warner~\cite{WarneRan65}, who introduced the RR mechanism for collecting sensitive data in social science research. RAPPOR~\cite{ErlinRAP14} extends RR to non-binary data by encoding values as $d$-bit vectors and applying RR to each bit. OLH~\cite{WangLoc17} refines this by introducing a \emph{local hash} to compress the $d$-bit vector, reducing communication cost. A comparative analysis of these mechanisms and their variants is in~\cite{CormodeFre21}.

Significant efforts have improved the privacy-utility tradeoff in LDP.
A variance analysis framework was proposed in~\cite{WangLoc17} to optimize the parameters of RR-based mechanisms, thereby enhancing data utility.
Post-processing techniques can also improve the utility.
For example, the non-negative and sum-to-one constraints were applied in~\cite{WangLocally20}, in which they referred to as \emph{consistency}. 
As another example, the convolution framework in~\cite{fang2023locally} added Wiener filter-based deconvolution to existing LDP protocols for improved data utility.  Interactive protocols such as PrivKV~\cite{YePri19} can iteratively improve estimation accuracy. Estimation of the most frequent items, or \emph{heavy hitters}, can be accomplished through random projection, as shown in~\cite{BassiLoc15}. Cryptographic methods enhance privacy without sacrificing utility, as seen in Crypt$\varepsilon$~\cite{RoyCry20}. However, none of them address correlated perturbation among contributors. Some techniques, including post-processing, can be integrated with JRR for further utility gains.

Privacy leakage due to data correlation has long been a concern. Prior research~\cite{KiferPuf14,YangBay15,SongPuf17,ChenCorr17,NiuUnl18} explores this from both theoretical and practical perspectives. The Pufferfish framework~\cite{KiferPuf14} enables customized privacy definitions for correlated data, later adapted in~\cite{SongPuf17}. Bayesian differential privacy~\cite{YangBay15} analyzes correlated data privacy from a Bayesian perspective, with~\cite{ChenCorr17} using Bayesian networks to determine the minimum required noise. A game-theoretic model~\cite{WuGam17} examines the privacy-utility tradeoff in data sharing. Applications such as graph data publication~\cite{LiGra16}, trajectory and network data release~\cite{ChenCorr17,OuRel18}, and trading statistics aggregation~\cite{NiuUnl18} have also been studied under differential privacy. However, these works focus on protecting correlated data, not the correlation among different contributors' perturbations.
 
The shuffle model~\cite{ErlinAmp19,BallePri19,BittaPro17,CheuDis19,MeehaPri22} enhances privacy by having a trusted auxiliary server shuffle perturbed data before forwarding it to the data collector, breaking the linkage between contributors and their data. Originally proposed in Prochlo~\cite{BittaPro17}, the model's theoretical privacy guarantees have since been extensively studied. The first instantiation of JRR also utilizes a non-colluding auxiliary server, but unlike the shuffle model, this server never accesses contributor-submitted data. Moreover, shuffling is complementary to JRR and can be integrated to further strengthen privacy.

A separate line of research focused on designing LDP mechanisms for various types of data,
including real-valued data~\cite{NguyeCol16, DuchiMin18,WangCol19,MaOPT21}, 
multi-dimensional data~\cite{WangCol19,XuCol20,RenLoP18,ChengTas21},  set-valued data~\cite{QinHea16,WangPri18,WangSet20}, time-series data~\cite{WangTow20}, 
social graph data~\cite{SunAna19}, key-value pairs~\cite{YePri19,GuPCK20,SunCon19}, sparse vector~\cite{zhou2022locally}, and directional data~\cite{WeggeDif21}. 
However, similar to existing LDP frequency estimation techniques, these works do not consider correlated perturbation. 

\section{Conclusions and Future Works}\label{Sec:Conl}
In this paper, we explored correlated random data perturbations for locally differentially private frequency estimation to achieve a better utility-privacy tradeoff. We have presented a general Joint Randomized Response (JRR) mechanism, along with two practical instantiations, which can provide the same level of data privacy as the classical RR mechanism while improving the data utility in an overwhelming majority of the cases. We have confirmed the advantages of JRR over RR through theoretical analysis and detailed simulation studies using both real and synthetic datasets. 

There are several directions to extend this work. First, since JRR may underperform RR if the ratio $n_1/n$ is very close to 0.5, it is possible to avoid this situation via a two-phase frequency estimation. In the first
phase, we use the standard RR to obtain a rough estimate of $n_1$ using a portion of privacy budget whereby to choose optimal $p$ and $\rho$ for JRR. In the second phase, we use JRR with these parameters to refine the estimation of $n_1$ using the remaining privacy budget. 
Additionally, we plan to extend JRR for groups with more than two contributors. Moreover, we will seek to extend JRR to support other data types such as non-binary data and explore its integration with advanced LDP mechanisms for other data analysis problems such as mean value estimation.

\begin{acks}
We would like to thank the anonymous reviewers for their insightful and constructive feedback, which has greatly helped us enhance the quality of our work. This work was supported in part by the US National Science Foundation under grants CNS-2245689 (CRII), CNS-2325564 and CNS-2325563, as well as the 2022 Meta Research Award for Privacy-Enhancing Technologies.
This work was also partially sponsored by the Army Research Laboratory and was accomplished under Cooperative Agreement Number W911NF-23-2-0225. The views and conclusions contained in this document are those of the authors and should not be interpreted as representing the official policies, either expressed or implied, of the Army Research Laboratory or the U.S. Government. The U.S. Government is authorized to reproduce and distribute reprints for Government purposes notwithstanding any copyright notation herein.
\end{acks}

\bibliographystyle{ACM-Reference-Format}
\bibliography{LDP_ref}

\appendix

\section{Proof of Theorem~\ref{thm:uEst_JRR}}\label{appendix:unbiased_jrr}

\begin{proof}
It is easy to see that the marginal probability distribution of $T_{2i-1}$ and $T_{2i}$ in Table.~\ref{tab:jointPro_2} is the same. 
More specifically, for any data contributor $u_j$, 
\begin{equation*}
T_j=\begin{cases}
1& \text{with probability $p$},\\
0& \text{with probability $q$}.\\
\end{cases}
\end{equation*}
Define $Y_j$ as the random indicator variable for data contributor $u_j$ reporting a perturbed value $y_j=1$.
There are two cases:
\begin{itemize}
    \item $x_j=0$, then $\mathrm{Pr}[Y_j=1|x_j=0]=\mathrm{Pr}[T_j=0]=q$;
    \item $x_j=1$, then $\mathrm{Pr}[Y_j=1|x_j=1]=\mathrm{Pr}[T_j=1]=p$.
\end{itemize} 

Denote $I_v$ as the random variable for the number of contributors reporting a perturbed value of $v$, where $v\in\{0,1\}$. 
When $v=1$, we have
\begin{equation*}
    I_1=\sum^{n}_{j=1}Y_j.
\end{equation*}
Taking the expectation on both sides follows 
\begin{equation}
\begin{split} 
    \mathrm{E}[I_1]&=\mathrm{E}[\sum^{n}_{j=1}Y_j]=\sum^{n}_{j=1}\mathrm{E}[Y_j]=\sum^{n}_{j=1}\mathrm{Pr}[Y_j=1]\\
    &=n_1\cdot \mathrm{Pr}[T_j=1]+(n-n_1)\cdot \mathrm{Pr}[T_j=0]\\
    &=n_1\cdot p+(n-n_1)\cdot q\\
    &=(p-q)n_1+nq.
\end{split}
\end{equation}
Therefore, the data collector can estimate the number of contributors having value $1$ as 
\begin{equation}
    \hat{n}_1=\frac{I_1-nq}{p-q},
\end{equation}
which is an unbiased estimator of $n_1$.

Similar to the proof of $\hat{n}_1$, we have $E[I_0]=(p-q)n_0+nq$ followed by $I_0=n-I_1$,
which leads to the same unbiased estimator in the theorem. 
\end{proof}

\section{Proof of Theorem~\ref{theo:privacy_level}}\label{appendix:jrr_privacy}
\begin{proof}
Denote by $\mathcal{C}$ the set of contributors who collude with the data collector and $\mathcal{T}_c=\{T_j|j\in\mathcal{C}\}$.  
We prove the theorem by showing
\begin{equation}
\begin{split}
        \frac{\Pr[\mathcal{M}(x_i)=y_i\ |\ \mathcal{T}_c]}{\Pr[\mathcal{M}(x'_i)=y_i\ |\ \mathcal{T}_c]} & \leq \frac{\max \Pr[\mathcal{M}(x_i)=y_i\ |\ \mathcal{T}_c]}{\min \Pr[\mathcal{M}(x'_i)=y_i\ |\ \mathcal{T}_c]} \\
    &=\frac{mp_{\max}+(n-m-1)p}{mp_{\min}+(n-m-1)q},
\end{split}
\end{equation}
for all $x_i,x'_i,y_i\in D$.

We start by analyzing $\Pr[\mathcal{M}(x_i)=y_i \ |\ \mathcal{T}_c]$. Specifically, denote by $u_j$ the contributor that is assigned to the same group as contributor $u_i$. We first have
\begin{equation}\label{eq:prob_yi}
    \begin{split} 
           &\Pr[\mathcal{M}(x_i)=y_i \ |\ \mathcal{T}_c]\\
        & =  \Pr[\mathcal{M}(x_i)=y_i \ |\ \mathcal{T}_c, j \notin \mathcal{C}] \cdot \Pr[j \notin \mathcal{C}]  \\
          & \quad + \Pr[\mathcal{M}(x_i)=y_i\ |\ \mathcal{T}_c, j \in \mathcal{C}] \cdot \Pr[j \in \mathcal{C}]    \\
    \end{split}
\end{equation}
Under uniform random grouping, we have 
\begin{equation} \label{eq:j_notin}
    \Pr[j \notin \mathcal{C}]=\frac{n-m-1}{n-1},
\end{equation}
and
\begin{equation} \label{eq:j_in}
  \Pr[j \in \mathcal{C}]=\frac{m}{n-1},
\end{equation}
where $m$ is the number of contributors who collude with the data collector. 

We now analyze the conditional probabilities of contributor $u_i$ reporting $y_i$ under these two cases.

\textbf{Case 1: $j \notin \mathcal{C}$}.
If $u_j$ is not a colluder, the probabilities of contributor $u_i$ reporting $y_i$ is independent of $\mathcal{T}_c$, and we have
\begin{equation}\label{eq:con_j_notin}
    \begin{split}
        \Pr[\mathcal{M}(x_i)=y_i \ |\ \mathcal{T}_c, j \notin \mathcal{C}] = \Pr[\mathcal{M}(x_i)=y_i].
    \end{split}
\end{equation}
It follows that
\begin{equation}\label{eq:con_j_notin_max}
    \begin{split}
         & \max_{x_i, y_i\in D} \Pr[\mathcal{M}(x_i)=y_i \ |\ \mathcal{T}_c, j \notin \mathcal{C}] \\
        = & \Pr[\mathcal{M}(x_i) = x_i]=p,
    \end{split}
\end{equation}
where the first equality means that the maximum is achieved when reporting truthfully (i.e., $y_i = x_i$). Similarly, we have
\begin{equation}\label{eq:con_j_notin_min}
    \begin{split}
          & \min_{x_i, y_i\in D} \Pr[\mathcal{M}(x_i)=y_i \ |\ \mathcal{T}_c, j \notin \mathcal{C}] \\
        = & \Pr[\mathcal{M}(x_i) = 1 - x_i]=q.
    \end{split}
\end{equation}


\textbf{Case 2: $j\in \mathcal{C}$}.
If $u_j$ colludes with the data collector, the conditional probabilities of contributor $u_i$ reporting $y_i$ only depend on $T_j$. We then have
\begin{equation}
    \begin{split}
        \Pr[\mathcal{M}(x_i)=y_i\ |\ \mathcal{T}_c, j \in \mathcal{C}] = \Pr[\mathcal{M}(x_i)=y_i \ |\ T_j].
    \end{split}
\end{equation}
There are four cases:
\begin{itemize}
    \item Case 2.1: If $T_i =1, T_j = 1$, then we have     \begin{equation}\label{eq:Gij=1_case1_T}
              \begin{split}
                     \Pr[\mathcal{M}&(x_i)=y_i \ |\ T_j] = \frac{\Pr[y_i=x_i,T_j=1]}{\Pr[T_j=1]} \\
                  = & \frac{p^2 + \rho pq}{p} = p + \rho q.
              \end{split}
          \end{equation}
    \item Case 2.2: If $T_i = 0, T_j = 1$, then we have \begin{equation}\label{eq:Gij=1_case2_T}
              \begin{split}
                     \Pr[\mathcal{M}&(x_i)=y_i \ |\ T_j] = \frac{\Pr[y_i= 1 - x_i,T_j=1]}{\Pr[T_j=1]} \\
                  = & \frac{(1 - \rho) pq}{p} = (1 - \rho) q.
              \end{split}
          \end{equation}
    \item Case 2.3: If $T_i = 1, T_j = 0$, then we have
          \begin{equation}\label{eq:Gij=1_case3_T}
              \begin{split}
                    \Pr[\mathcal{M}&(x_i)=y_i \ |\ T_j] = \frac{\Pr[y_i= x_i,T_j=0]}{\Pr[T_j=0]} \\
                  = & \frac{(1 - \rho) pq}{q} = (1 - \rho) p.
              \end{split}
          \end{equation}
    \item Case 2.4: If $T_i = 0, T_j = 0$, then we have
          \begin{equation}\label{eq:Gij=1_case4_T}
              \begin{split}
                    \Pr[\mathcal{M}&(x_i)=y_i \ |\ T_j] = \frac{\Pr[y_i= 1 - x_i,T_j=0]}{\Pr[T_j=0]} \\
                  = & \frac{q^2 + \rho pq}{q} = q + \rho p.
              \end{split}
          \end{equation}
\end{itemize}
The maximum and the minimum of the above four cases are given by
\begin{equation}\label{eq:con_kin_gik1_max}
    \begin{split}
        p_{\max} & = \max\{p + \rho q, (1 - \rho) q, (1 - \rho) p, q + \rho p\} \\
                 & = \max\{p + \rho q, (1 - \rho) p\},                          \\
        p_{\min} & = \min\{p + \rho q, (1 - \rho) q, (1 - \rho) p, q + \rho p\} \\
                 & = \min\{q + \rho p, (1 - \rho) q\}.
    \end{split}
\end{equation}
It follows that
\begin{equation}\label{eq:con_j_in_max}
    \begin{split}
          \max_{x_i, y_i\in D} \Pr[\mathcal{M}(x_i)=y_i | \mathcal{T}_c, j \in \mathcal{C}]= p_{\max}
    \end{split}
\end{equation}
and
\begin{equation}\label{eq:con_j_in_min}
    \begin{split}
        \min_{x_i, y_i\in D} \Pr[\mathcal{M}(x_i)=y_i | \mathcal{T}_c, j \in \mathcal{C}]= p_{\min}
    \end{split}
\end{equation}
Substituting Eqs.~ (\ref{eq:j_notin}), (\ref{eq:j_in}),(\ref{eq:con_j_notin_max}) and (\ref{eq:con_j_in_max}) into Eq.~(\ref{eq:prob_yi}), we have
\begin{equation}\label{eq:yj_max}
    \begin{split}
        \max_{x_i, y_i} \Pr[\mathcal{M}(x_i)=y_i|\mathcal{T}_c]
        = \frac{mp_{\max}}{n - 1} + \frac{n - 1 - m}{n - 1} \cdot p,
    \end{split}
\end{equation}
Similarly, substituting Eqs.~(\ref{eq:j_notin}), (\ref{eq:j_in}),  (\ref{eq:con_j_notin_min}) and (\ref{eq:con_j_in_min}) into Eq.~(\ref{eq:prob_yi}), we have
\begin{equation}\label{eq:yj_min}
    \begin{split}
          \min_{x_i, y_i} \Pr[\mathcal{M}(x_i)=y_i|\mathcal{T}_c]
      =  \frac{mp_{\min} }{n - 1}+ \frac{n - 1 - m}{n - 1} \cdot q,
    \end{split}
\end{equation}
It follows that
\begin{equation}
    \begin{split}
        \frac{\max \Pr[\mathcal{M}(x_i)=y_i\ |\ \mathcal{T}_c]}{\min \Pr[\mathcal{M}(x'_i)=y_i\ |\ \mathcal{T}_c]} 
         =
        \frac{mp_{\max}+(n-m-1)p}{mp_{\min}+(n-m-1)q}\;,
    \end{split}
\end{equation}
for all $x_i,x'_i,y_i\in D$.
The theorem is thus proved.
\end{proof}

\section{Proof of Theorem~\ref{thm:variance}} \label{appendix:jrr_utility}


\begin{proof}
First, 
    the variance of the estimator $\hat{n}_v$ is given by
\begin{equation}\label{eq:var_nv}
    \mathrm{Var}[\hat{n}_v]=\frac{\mathrm{Var}[I_v-nq]}{(p-q)^2}=\frac{\mathrm{Var}[I_v]}{(p-q)^2},
\end{equation}
where the second equality holds because both $n$ and $q$ are constant. 
Since $n = n_0 + n_1$, we have $\mathrm{Var}[\hat{n}_0]=\mathrm{Var}[n-\hat{n}_1]=\mathrm{Var}[\hat{n}_1]$. In what follows, we focus on the analysis of $\mathrm{Var}[\hat{n}_1]$. 

Again define $Y_j$ to be the indicator random variable such that $Y_j=1$ if contributor $u_j$ reports a perturbed value of ``1'' and $0$ otherwise for all $1\leq j\leq n$. 
Without loss of generality, assume that group $G_i$ consists of contributors $u_{2i-1}$ and $u_{2i}$ for all $1\leq i\leq n/2$.  
Since the perturbation of different groups is independent of each other, we have
\begin{equation}\label{equ:variance_of_each_group}
    \begin{split}
        \mathrm{Var}[I_1] =\mathrm{Var}[\sum^{n}_{j=1}Y_j]=\sum^{n/2}_{i=1}\mathrm{Var}[Y_{2i-1}+Y_{2i}].
    \end{split}
\end{equation}

The $n/2$ groups can be classified into three categories: Type-1 groups with both contributors having value $1$, Type-2 group with one contributor having value $1$ and the other having value $0$, and Type-3 groups with both contributors having value $0$. 
The variance of each group's variance $\mathrm{Var}[Y_{2i-1}+Y_{2i}]$ depends on its type, and groups of the same type have the same variance. 
Define $V_z=\mathrm{Var}[Y_{2i-1}+Y_{2i}]$ if group $G_i$ is a type-$z$ group for all  $1\leq z\leq 3$ and $1\leq i\leq \frac{n}{2}$.
Let $m_{1}$, $m_2$, and $m_3$ be the numbers of Type-1, Type-2, and Type-3 groups, respectively, which are themselves random variables due to uniform random grouping. For any given $n_0$ and $n_1$, we have $2m_1+m_2=n_1$ and
$m_1+m_2+m_3=n/2$. It follows that $m_2=n_1-2m_1$ and $m_3=m_1+n/2-n_1$, which indicates that the random grouping only produces one independent random variable $m_1$. 

For any given $m_1$, the conditional variance of $I_1$ is given by 
\begin{equation}\label{eq:var_I1_m1}
\begin{split}
        \mathrm{Var}&[I_1|m_1]=\sum^{n/2}_{i=1}\mathrm{Var}[Y_{2i-1}+Y_{2i}]\\ 
        =&m_{1}V_{1} + m_{2}V_{2} + m_{3}V_{3}\\
       = &m_{1}V_{1} + (n_1 - 2m_{1})V_{2} + (m_{1} + \frac{n}{2}- n_1)V_{3}.
\end{split}\end{equation}
According to the law of total variance \cite{LawVar}, the (unconditional) variance of $I_1$ is given by 
\begin{equation}\label{equ:total_var}
       \mathrm{Var}[I_1] 
       = \mathrm{E}[\mathrm{Var}[I_1|m_{1}]] + \mathrm{Var}[\mathrm{E}[I_1|m_{1}]].
\end{equation}

Next, we calculate the two terms in Eq.~(\ref{equ:total_var}) one by one.

\textbf{The first term \boldmath{$\mathrm{E}[\mathrm{Var}[I_1|m_{1}]]$}.} 
We first calculate $V_1$, $V_2$ and $V_3$. For any group $G_i$, we have
\begin{equation}\label{eq:Var12}
    \begin{split}
        & \mathrm{Var}[Y_{2i-1}+Y_{2i}] \\
        = & \mathrm{Var}[Y_{2i-1}] + \mathrm{Var}[Y_{2i}] + 2\mathrm{Cov}[Y_{2i-1}, Y_{2i}] \\
        = & \mathrm{Var}[Y_{2i-1}] + \mathrm{Var}[Y_{2i}] \\
        &+ 2(\mathrm{E}[Y_{2i-1}Y_{2i}]-\mathrm{E}[Y_{2i-1}]\mathrm{E}[Y_{2i}]).
    \end{split}
\end{equation}
There are three cases.
\begin{itemize}
\item Case~1: If $G_i$ is of Type-1, then we have
\begin{equation}\label{eq:case1.1}
\mathrm{E}[Y_{2i-1}Y_{2i}]=\mathrm{Pr}[T_{2i-1}=1,T_{2i}=1]=\rho p q+p^2,
\end{equation}
and
\begin{equation}\label{eq:case1.2}
    \mathrm{E}[Y_{2i-1}]\mathrm{E}[Y_{2i}]=\mathrm{Pr}[T_{2i-1}=1]\cdot \mathrm{Pr}[T_{2i}=1]=p^2.
\end{equation}
\item Case~2: If $G_i$ is of Type-2, then we have
\begin{equation}\label{eq:case2.1}
    \mathrm{E}[Y_{2i-1}Y_{2i}]=(1-\rho)pq
\end{equation}
and
\begin{equation}\label{eq:case2.2}
    \mathrm{E}[Y_{2i-1}]\mathrm{E}[Y_{2i}]=pq.
\end{equation}
\item Case~3: If $G_i$ is of Type-3, then we have
\begin{equation}\label{eq:case3.1}
    \mathrm{E}[Y_{2i-1}Y_{2i}]=q^2+\rho pq,
\end{equation}
and
\begin{equation}\label{eq:case3.2}
    \mathrm{E}[Y_{2i-1}]E[Y_{2i}]=q^2.
\end{equation}
\end{itemize}
 Substituting Eqs.~(\ref{eq:case1.1}) to (\ref{eq:case3.2}) into Eq.~(\ref{eq:Var12}), we get
\begin{equation}\label{eq:V123_gen}
    \begin{split}
    V_1&=2pq(1+\rho),\\
    V_2&=2pq(1-\rho),\\
    V_3&=2pq(1+\rho).
\end{split}
\end{equation}
Substituting Eq.~(\ref{eq:V123_gen}) into Eq.~(\ref{eq:var_I1_m1}), we have
\begin{equation}
    \begin{split}
        \mathrm{Var}[I_1|m_{1}]
        = npq + (8m_{1} + n - 4n_1)\rho pq,
    \end{split}
\end{equation}
Taking the expectation on both sides, we have
\begin{equation}\label{eq:E_Con_var}
    \begin{split}
        \mathrm{E}[\mathrm{Var}[I_1|m_{1}]] 
        = & \mathrm{E}[npq + (8m_{1} + n - 4n_1)\rho pq] \\
        = & npq + (8\mathrm{E}[m_{1}] + n - 4n_1)\rho pq.
    \end{split}
\end{equation}
Since the expectation of the number of Type-1 groups is 
\begin{equation}\label{eq:E_m1}
    \mathrm{E}[m_{1}] = \frac{n}{2}\cdot \frac{n_1(n_1-1)}{n(n-1)}= \frac{n_1(n_1-1)}{2(n-1)},
\end{equation}
Substituting Eq.~(\ref{eq:E_m1}) into Eq.~(\ref{eq:E_Con_var}), we have
\begin{equation} \label{eq:Evar_I1_m1}
    \mathrm{E}[\mathrm{Var}[I_1|m_{1}]] = npq + \frac{(2n_1 - n)^2 - n}{n-1}\rho pq.
\end{equation}

\textbf{The second term \boldmath{$\mathrm{Var}[\mathrm{E}[I_1|m_{1}]]$}.} 
According to the definition of conditional expectation, we have
\begin{equation}
\begin{split}
        \mathrm{E}[I_1|m_{1}]&=\mathrm{E}[\sum_{j=1}^n Y_j|m_1]=\sum_{j=1}^n \mathrm{E}[Y_j|m_1]\\
        &=n\cdot 1\cdot \Pr(Y_j=1|m_1).
\end{split}
\end{equation}
Under JRR, whether an arbitrary contributor $u_j$ reports $1$ or 0 only depends on the contributor's original value $x_j$ and the identical marginal probability distribution $\mathrm{Pr}[T_j]$. Since the numbers of contributors with the original value 1 and 0, $n_1$ and $n_0$, are predetermined. Thus, we have 
\begin{equation}
\begin{split}
        &\mathrm{E}[I_1|m_1]=n\cdot \Pr[Y_j=1|m_1]\\
        &\quad=n_1\cdot \Pr[T_j=1]+ (n-n_1)\cdot \Pr[T_j=0]\\
         &\quad=n_1p+ (n-n_1)(1-p)\\
        &\quad =(2n_1 - n)p + n - n_1,
\end{split}
\end{equation}
which is a constant independent with $m_1$. It follows that
\begin{equation}\label{eq:varE_I1_m1}
    \mathrm{Var}[\mathrm{E}[I_1|m_{1}]]=0.
\end{equation}

Substituting Eqs.~(\ref{eq:varE_I1_m1}) and (\ref{eq:Evar_I1_m1}) into Eq.~(\ref{equ:total_var} ), we have
\begin{equation} \label{eq:var}
    \mathrm{Var}[I_1] = npq + \frac{(2n_1 - n)^2 - n}{n-1}\rho pq.
\end{equation}
  
Finally, substituting Eq.~(\ref{eq:var}) into Eq.~(\ref{eq:var_nv}), we have
\begin{equation*}
    \mathrm{Var}[\hat{n}_v] = \frac{pq}{(p-q)^2} \cdot (n+\frac{\rho((2n_1-n)^2-n)}{n-1}).
\end{equation*}
The theorem is thus proved.
\end{proof}

\section{Proof of the Lemma~\ref{lemma:1}}\label{appendix:lemma:1}
\begin{proof}
Let $(2n_1-n)^2-n<0$. Solving the inequality, we have
\begin{equation}
-\sqrt{n}<2n_1-n<\sqrt{n}.
\end{equation}
By simple algebraic manipulation, we get
\begin{equation}
\frac{n-\sqrt{n}}{2}<n_1<\frac{n-\sqrt{n}}{2}.
\end{equation}
Dividing all three sides by $n$, we can obtain
\[
\frac{n-\sqrt{n}}{2n}<\frac{n_1}{n}<\frac{n-\sqrt{n}}{2n}.
\]
We therefore have $(2n_1-n)^2-n<0$ if
$
\frac{1}{2}-\frac{1}{2\sqrt{n}}<\frac{n_1}{n}<\frac{1}{2}+\frac{1}{2\sqrt{n}}$
and $(2n_1-n)^2-n\geq 0$  if $\frac{n_1}{n} \in [0,\frac{1}{2}-\frac{1}{2\sqrt{n}}]\bigcup [\frac{1}{2}+\frac{1}{2\sqrt{n}}]$
The lemma is thus proved.
\end{proof}

\section{Proof of  Lemma~\ref{lemma:2}}\label{appendix:lemma:2}
\begin{proof}
Since $0\leq n_1\leq n$, we have $0\leq (2n_1-n)^2\leq n^2$. Subtracting $n$ from all three sides and then dividing them by $n-1$, we get
\begin{equation}
\frac{0-n}{n-1}\leq \frac{(2n_1-n)^2-n}{n-1}\leq \frac{n^2-n}{n-1}.
\end{equation}
It follows that
\begin{equation}\label{eq:3}
\frac{1}{n-1}-1\leq \frac{(2n_1-n)^2-n}{n-1}\leq n.
\end{equation}
Since $\rho\in[-1,1]$ and $n\geq 2>|\frac{1}{n-1}-1|$, multiplying $\rho$ by all three sides of Inequality~(\ref{eq:3}), we get
\begin{equation}
-n<\rho(\frac{1}{n-1}-1)\leq \rho\cdot \frac{(2n_1-n)^2-n}{n-1}\leq \rho n\leq n.
\end{equation}
It follows that 
\begin{equation}\label{eq:4}
-n< \rho\cdot \frac{(2n_1-n)^2-n}{n-1}\leq n.
\end{equation}
Adding $n$ to all three sides of Inequality~(\ref{eq:4}), we get
\begin{equation}
0< n+\rho\cdot \frac{(2n_1-n)^2-n}{n-1}\leq 2n.
\end{equation}
It follows that 
\begin{equation}
n+\rho\cdot \frac{(2n_1-n)^2-n}{n-1}> 0.
\end{equation}
The lemma is thus proved.

\end{proof}

\section{Proof of  Lemma~\ref{lemma:3}}\label{appendix:lemma:3}
\begin{proof}
Since $q=1-p$, we have 
\[
\frac{pq}{(p-q)^2}=\frac{p(1-p)}{(2p-1)^2}= \frac{1}{4}(\frac{1}{(2p-1)^2}-1).
\]
It is easy to see that $\frac{1}{(2p-1)^2}$ is monotonically decreasing with respect to $p\in (0.5,1]$. Therefore, $\frac{pq}{(p-q)^2}$ is also monotonically decreasing with respect to $p\in (0.5,1]$. The lemma is therefore proved.
\end{proof}

\section{Proof of  Theorem~\ref{thm:mono}}\label{appendix:thm:mono}
\begin{proof}
Since $h(p,\rho)$ is the product of $\frac{pq}{(p-q)^2}$ and $n+\frac{\rho((2n_1-n)^2-n)}{n-1}$ according to Eq.~(\ref{eq:variance_2u}), we can analyze its monotonicity with respect to $p$ and $\rho$ based on the monotonicity of $\frac{pq}{(p-q)^2}$ and $n+\frac{\rho((2n_1-n)^2-n)}{n-1}$.

First, since $\frac{pq}{(p-q)^2}$ is monotonically decreasing with respect to $p\in(0.5,1]$ according to Lemma~\ref{lemma:3}, and $n+\frac{\rho((2n_1-n)^2-n)}{n-1}>0$ according to Lemma~1 and is independent of $p$, $h(p,\rho)$ is monotonically decreasing with respect to $p\in(0.5,1]$. 

Second, since $\frac{pq}{(p-q)^2}>0$ and is independent of $\rho$, the monotonicity of $h(p,\rho)$ with respect to $\rho$ is the same as that of $n+\frac{\rho((2n_1-n)^2-n)}{n-1}$. Since $n\geq 2$ and $((2n_1-n)^2-n)<0 $ if $n_1/n\in (\frac{1}{2}-\frac{1}{2\sqrt{n}},\frac{1}{2}+\frac{1}{2\sqrt{n}})$
according to Lemma~\ref{lemma:2}, $h(p,\rho)$ is also  monotonically decreasing with respect to $\rho$ if $\frac{n_1}{n}\in (\frac{1}{2}-\frac{1}{2\sqrt{n}},\frac{1}{2}+\frac{1}{2\sqrt{n}})$. By similar deduction, it is also easy to prove that $h(p,\rho)$ is also monotonically increasing with respect to $\rho$ if $\frac{n_1}{n} \in [0,\frac{1}{2}-\frac{1}{2\sqrt{n}}]\bigcup [\frac{1}{2}+\frac{1}{2\sqrt{n}},1]$. 

The theorem is therefore proved.
\end{proof}

\section{Monotonicity of $f(m)$} \label{appendix:proof:m_M}

Denote by $g_1(m)=mp_{\max}+(n-m-1)p$ and $g_2(m)=mp_{\min}+(n-m-1)q$. We have $f(m)= \frac{g_1(m)}{g_2(m)}$, and its derivative is
\begin{equation}\label{eq:diff}
\begin{split}
    f'(m)&=\frac{g'_1(m)\cdot g_2(m)-g_1(m)\cdot g'_2(m)}{g^2_2(m)}\\
    &=\frac{(p_{\max}\cdot q-p_{\min}\cdot p)\cdot (n-1)}{g^2_2(m)} 
    \end{split}
\end{equation}

We now consider the following two cases.
\begin{itemize}
    \item Case 1: if $\rho\leq 0$, we have $p_{\max}=(1-\rho)p$ and $p_{\min}=q+\rho p$. It follows that 
    $p_{\max}\cdot q-p_{\min}\cdot p=-2\rho pq >0$,
    \item Case 2: if $\rho>0$, we have $p_{\max}=p+\rho q$ and $p_{\min}=(1-\rho)q$. 
    It follows that $p_{\max}\cdot q-p_{\min}\cdot p=2\rho pq >0$.
\end{itemize}
Notice that $g^2_2(m)>0$ and $n-1>0$. We then have $f'(m)>0$, and $f(m)$ is monotonically increasing with respect to $m$. 

\section{Proof of Theorem~\ref{thm:practicalM}}\label{appendix:proof:practicalM}
 \begin{proof}
    We prove the reporting trustfulness in Section~\ref{Sec:Practice} is the same as in Table~\ref{tab:jointPro_2}.

For any group with two contributors $u_{2i-1}$ and $u_{2i}$, let $T_{2i-1}$ and $T_{2i}$ be the truthfulness of the two contributors' reports. 

First, for the case $T_{2i-1}=1,T_{2i}=1$, we have:
\begin{equation}
    \begin{split}
        &\mathrm{Pr}[T_{2i-1}=1,T_{2i}=1] \\
        =&\mathrm{Pr}[T_{2i-1}=1,T_{2i}=1| R_1=1] \\
        &+ \mathrm{Pr}[T_{2i-1}=1,T_{2i}=1| R_1=-1] \\
        =&\frac{1}{2}(p^2 + \rho pq) +\frac{1}{2}(p^2 + \rho pq)
        =p^2 + \rho pq.
    \end{split}
\end{equation}

Second, for the case $T_{2i-1}=1,T_{2i}=0$, we have:
    \begin{equation}
        \begin{split}
            &\mathrm{Pr}[T_{2i-1}=1,T_{2i}=0] \\
            =&\mathrm{Pr}[T_{2i-1}=1,T_{2i}=0| R_1=1] \\
            &+ \mathrm{Pr}[T_{2i-1}=1,T_{2i}=0| R_1=-1] \\
            =&\frac{1}{2}((1-\rho)pq - \sqrt{-\rho pq}) \\
            &+\frac{1}{2}((1-\rho)pq + \sqrt{-\rho pq})\\
            =&(1-\rho)pq.
        \end{split}
    \end{equation}
    $T_{2i-1}=0,T_{2i}=1$ is symmetric to the case of $T_{2i-1}=1,T_{2i}=0$,
    so we have $\mathrm{Pr}[T_{2i-1}=0,T_{2i}=1]=(1-\rho)pq$.
   
    For the case of $T_{2i-1}=0,T_{2i}=0$, we have
    \begin{equation}
        \begin{split}
            &\mathrm{Pr}[T_{2i-1}=0,T_{2i}=0] \\
            =&\mathrm{Pr}[T_{2i-1}=0,T_{2i}=0| R_1=1] \\
            &+ \mathrm{Pr}[T_{2i-1}=0,T_{2i}=0| R_1=-1] \\
            =&\frac{1}{2}(q^2 - \rho pq) \\
            &+\frac{1}{2}(q^2 - \rho pq)\\
            =&q^2 -\rho pq.
        \end{split}
    \end{equation}
    These results are the same as in Table~\ref{tab:jointPro_2}.
\end{proof}

\section{Details of Real-world Datasets} \label{appendix:datasets}
We use the following four real-world datasets to evaluate the performance of JRR:
\begin{itemize}
\item \textbf{Kosarak \cite{Kosarak}}: a dataset containing the click stream of a Hungarian news website that records about 8 million click events for $41,270$ different pages. For our purpose, we randomly select $100$ pages as the target pages and $20,000$ click events as contributors. If a click event's visited page belongs to the target pages, that contributor's true value is \emph{``1: visited'' } and \emph{``0: not'' } otherwise. The frequency of the clicks on the target pages is deemed as the ground truth.
\item \textbf{Amazon Rating Dataset \cite{Amazon}:} a dataset that contains over $2$ million customer ratings of beauty-related products sold on Amazon. We randomly select $10,000$ customers as contributors and set each contributor's true value to \emph{1} if his/her rating is ``1 star'' and \emph{0} otherwise.
\item \textbf{E-commerce \cite{E-Commerce}:} a women's clothing E-Commerce dataset consisting of $23,486$ records and $10$ features variables. We select the binary variable ``Recommended IND'' as each contributor's true data.
\item \textbf{Census\cite{census}:} a dataset of the United States census in 2010 from the Integrated Public Use Microdata Series (IPUMS). We randomly select $10,000$ records and set each contributor's true value to \emph{1} if the code of group quarter (GQ) is 1 and \emph{0} otherwise.
\end{itemize}

\section{Marginal distribution and estimator of $k$-JRR} \label{appendix:proof:extension}
We first prove that the marginal distribution of $k$-JRR (Section~\ref{sec:extension-non-binary}) is identical, with each contributor reporting their true value with probability $p$ and any other value with probability $q$.
\begin{proof}
    Let $v_1$ and $v'_1$ be a contributor's true value and reported value. We have 
    \begin{equation}
        \begin{split}
            &\Pr[v_1' = v_1] = \sum_{v_2'\in [k]} \Pr[v_1' = v_1, v_2'] \\
            =& p^2 + \rho pq + (k-1)(pq - \frac{1}{k-1}\rho pq) \\ 
            =& p.
        \end{split}
    \end{equation}
    Similarly, for each $v_1' \neq v_1$ in the data domain, we have
    \begin{equation}
        \begin{split}
            &\Pr[v_1' \neq v_1] = \sum_{v_2'\in [k]} \Pr[v_1' \neq v_1, v_2'] \\
            =& (pq - \frac{1}{k-1}\rho pq) + (k-1)(q^2 + \frac{1}{(k-1)^2}\rho pq) \\ 
            =& q.
        \end{split}
    \end{equation}

   We now prove the estimator $\hat{n}_v = (I_v - nq)/(p - q)$ is unbiased. First, we have
    \begin{equation}
        \begin{split}
            & \mathrm{E}[I_v] = n_v \cdot \Pr[v' = v] + (n - n_v) \cdot \Pr[v' \neq v] \\
            =& n_v p + (n - n_v)q.
        \end{split}
    \end{equation}
    Plugging $\mathrm{E}[I_v]$ into $\hat{n}_v$ gives 
    \begin{equation}
        \begin{split}
            &\mathrm{E}[\hat{n}_v] = \frac{\mathrm{E}[I_v] - n q}{p - q} \\
            =& \frac{n_v p + (n - n_v)q - n q}{p - q} \\
            =& n_v,
        \end{split}
    \end{equation}
    i.e. $\hat{n}_v$ is an unbiased estimator for $n_v$.
\end{proof}

\end{document}